\documentclass[times,twocolumn,final,authoryear]{elsarticle}
\usepackage{graphicx,stfloats,times,commath,amssymb,amsmath,amsfonts,subfigure,url,multirow,bm,algorithm,algorithmic,color,booktabs}
\usepackage[flushleft]{threeparttable}
\usepackage{multirow,bm,multicol,booktabs,extpfeil}



\journal{ }
\begin{document}

\begin{frontmatter}

\title{Channel Reflection: Knowledge-Driven Data Augmentation\\ for EEG-Based Brain-Computer Interfaces}
\author[1,2]{Ziwei Wang\corref{cor1}}\ead{vivi@hust.edu.cn}
\author[1,2]{Siyang Li\corref{cor1}}\ead{syoungli@hust.edu.cn}
\author[3]{Jingwei Luo}\ead{jonnylaw@163.com}
\author[4]{Jiajing Liu}\ead{liu_jiajing@hust.edu.cn}
\author[1,2]{Dongrui Wu\corref{cor2}}\ead{drwu@hust.edu.cn}

\address[1]{Key Laboratory of the Ministry of Education for Image Processing and Intelligent Control, School of Artificial Intelligence and Automation, Huazhong University of Science and Technology, Wuhan 430074, China}
\address[2]{Shenzhen Huazhong University of Science and Technology Research Institute, Shenzhen 518063, China}
\address[3]{China Electronic System Technology Co., Ltd., Beijing 100089, China}
\address[4]{School of Civil and Hydraulic Engineering, Huazhong University of Science and Technology, Wuhan 430074, China}

\begin{abstract}
A brain-computer interface (BCI) enables direct communication between the human brain and external devices. Electroencephalography (EEG) based BCIs are currently the most popular for able-bodied users. To increase user-friendliness, usually a small amount of user-specific EEG data are used for calibration, which may not be enough to develop a pure data-driven decoding model. To cope with this typical calibration data shortage challenge in EEG-based BCIs, this paper proposes a parameter-free channel reflection (CR) data augmentation approach that incorporates prior knowledge on the channel distributions of different BCI paradigms in data augmentation. Experiments on eight public EEG datasets across four different BCI paradigms (motor imagery, steady-state visual evoked potential, P300, and seizure classifications) using different decoding algorithms demonstrated that: 1) CR is effective, i.e., it can noticeably improve the classification accuracy; 2) CR is robust, i.e., it consistently outperforms existing data augmentation approaches in the literature; and, 3) CR is flexible, i.e., it can be combined with other data augmentation approaches to further increase the performance. We suggest that data augmentation approaches like CR should be an essential step in EEG-based BCIs. Our code is available online.
\end{abstract}

\end{frontmatter}

\section{Introduction}

A brain-computer interface (BCI) enables direct communication between the human brain and an external device \citep{Rosenfeld2017}. BCIs serve diverse purposes, encompassing research, mapping, augmentation, assistance, and restoration of human cognitive and/or sensory-motor functions \citep{Krucoff2016}.

According to the proximity of the electrodes to the brain cortex, BCIs can be categorized into three types: non-invasive, partially invasive, and invasive \citep{Wu2020}. The latter two require the surgical placement of sensors, and hence are predominantly employed in clinical applications. Non-invasive BCIs are the preferred choice for able-bodied individuals. They could use different input signals, e.g., electroencephalography (EEG), magnetoencephalography, functional magnetic resonance imaging, and functional near-infrared spectroscopy. Among them, EEG stands as the most popular choice due to its convenience and cost-effectiveness. Classical paradigms of EEG-based non-invasive BCIs include motor imagery (MI) \citep{Pfurtscheller2001}, steady-state visual evoked potential (SSVEP) \citep{Friman2007}, and event-related potential (ERP) \citep{Hoffmann2008}. Additionally, BCIs have also been used for epileptic seizure detection \citep{Acharya2013}, driver-drowsiness estimation \citep{Lin2005}, emotion analysis \citep{Wu2023}, and so on.

\begin{figure*}[htpb] \centering
\includegraphics[width=\linewidth,clip]{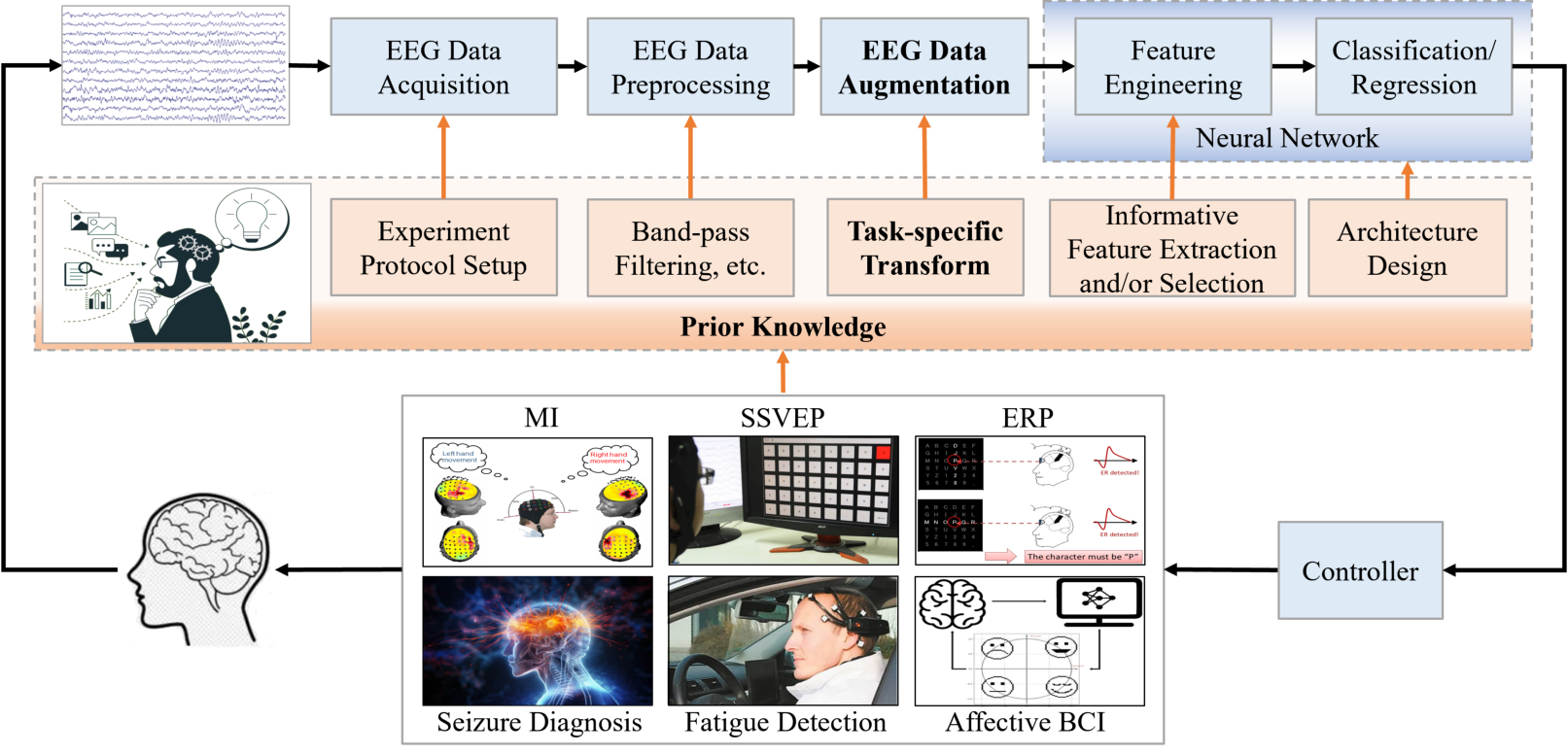}
\caption{Knowledge-driven machine learning pipeline for EEG-based BCIs, which includes EEG data acquisition, data preprocessing, data augmentation (optional), feature engineering, and classification/regression. The latter two can be integrated into a single end-to-end neural network. Task-related prior knowledge can be integrated into all blocks of the pipeline.} \label{fig:Knowledge_BCI}
\end{figure*}

The development of BCIs heavily relies on comprehending how the human brain functions. The `homunculus' model from the 20th century confirmed the presumed relationship between each part of the human body and a corresponding region in the primary motor/somatosensory areas of the neocortex \citep{Jasper1949}, which becomes the basis for MI-based BCIs \citep{Pfurtscheller2001}. Through differentiating the patterns in brain signals of imaged movements of different body parts, MI-based BCIs can assist, augment, or even repair human sensory-motor functions in various applications \citep{Krucoff2016}. The mechanism relies on the neuroscience discovery that processing of motor commands or somatosensory stimuli causes an attenuation of the rhythmic activity termed event-related desynchronization, while an increase is termed as event-related synchronization \citep{Blankertz2008}.

Similarly, other paradigms also follow specific neuroscience discoveries. SSVEPs are electrical brain responses that occur in synchrony with repetitive visual stimuli, such as flashing lights or flickering images, typically elicited in the visual cortex of the brain, specifically in the occipital region located at the back of the head \citep{Friman2007}. ERPs are triggered by specific events or stimuli, offering valuable insights into cognitive processes, such as attention, memory, and perception. The P300 ERP is distinguished by a prominent positive deflection within the EEG signal, typically $\sim$300 milliseconds after the presentation of a sensory stimulus \citep{Hoffmann2008}. The P300 ERP is prominently associated with neural activity in the parietal cortex, particularly in the parietal midline (centroparietal) region \citep{Abiri2019}.

EEG-based seizure classification emerges as a pivotal focus within the field of neurology, with positive implications for patients affected by epilepsy \citep{Acharya2013}. Typically, seizures can originate in various brain regions, including the temporal, frontal, parietal and occipital lobes. Seizure activity can spread from the onset zone to other brain parts. The path and extent of propagation can vary, leading to different types of seizures and clinical manifestations.

Accurate brain signal decoding is critical for successful BCI applications. Although EEG-based BCIs have various advantages as discussed, there are still many challenges for their wide-spread real-world applications \citep{lance2012brain,makeig2012evolving}, including non-stationarity of EEG signals, individual differences, and inter-environment differences. The last refers to the differences in EEG headsets and experiment protocols across datasets. Such differences limit the current analysis of EEG within a given paradigm and dataset, limiting the number of training samples.

Data augmentation is the most commonly used strategy to cope with the small data problem \citep{Shorten2019,LI2022danlp}. For EEG signal analysis, approaches from both signal processing and deep learning have been attempted. Nevertheless, such approaches are usually based on the characteristics of time series, EEG signals, or neural networks in general. Without adequately utilizing the paradigm-specific knowledge, such approaches usually have unstable performance and are not generalizable across paradigms and datasets. A simple, effective and generalizable data augmentation approach remains to be found. Notably, the relationships among paradigms are complex yet vital. Delving into the connections among channels, especially in different brain regions, is essential for comprehending the brain and contributing to constructing machine learning models.

The role of prior knowledge has received much attention, particularly for interpretable and explainable models \citep{lisboa2023coming}. \citet{Rueden2021} pointed out that ``\emph{Despite its great success, machine learning can have its limits when dealing with insufficient training data. A potential solution is the additional integration of prior knowledge into the training process}." In EEG-based BCIs, solely data-driven machine learning approaches are becoming the current trend for signal decoding. Nevertheless, effective integration of prior knowledge can profoundly influence model performance at various stages, as illustrated in Figure~\ref{fig:Knowledge_BCI}.

This paper proposes a straightforward yet effective knowledge-driven channel reflection (CR) data augmentation approach to generate high-quality task-specific augmented data, enriching the training dataset without introducing additional hyperparameters. Inspired by the left/right hand MI paradigm design, CR constructs new samples by reflecting left and right brain electrodes/channels of EEG signals and simultaneously exchanging the labels. For other BCI paradigms, CR assumes task-invariability when reflecting the hemispheres and sticks with the original labels. Extensive experiments on eight public EEG datasets demonstrated that CR could effectively improve performance on four BCI paradigms, namely MI, SSVEP, P300, and seizure classification. Our Python code is available on GitHub\footnote{https://github.com/sylyoung/DeepTransferEEG}.

The remainder of this paper is organized as follows: Section~\ref{sect:relatedwork} introduces related works. Section~\ref{sect:approach} proposes CR. Section~\ref{sect:experiments} presents experimental results to show the effectiveness of CR. Finally, Section~\ref{sect:conclusions} draws conclusions and points out future research directions.

\section{Related Works}\label{sect:relatedwork}

This section introduces related works on integrating prior knowledge into machine learning in EEG-based BCIs, and discusses current data augmentation approaches for EEG classification.

\subsection{Integration of Prior Knowledge in EEG-based BCIs}

Previous works have demonstrated the power of prior knowledge in various stages in brain signal decoding:
\begin{enumerate}
\item Data Preprocessing. Filtering and denoising procedures \citep{pedroni2019preprocess}, vital for data quality enhancement, depend extensively on the specific characteristics and requirements of the dataset. Prior knowledge aids in the development of robust data preprocessing methods, ensuring that the data fed into the model is clean, reliable, and aligned with the objectives of the task. As an example, the passband of filters differs across BCI paradigms.

\item Feature Engineering. Features can be extracted in the time domain \citep{boonyakitanont2020review}, frequency domain, time-frequency domain, etc \citep{Jenke2014}. However, without integration of task specific knowledge, general features for time series would not work well for a particular BCI paradigm. For example, in EEG-based BCIs, selecting features or spatial patterns relies on prior knowledge of neurophysiological processes and specific cognitive tasks. For MI-based BCIs, Common Spatial Pattern (CSP) \citep{Ramoser2000} is the most widely used supervised spatial filter. It aims to find a set of spatial filters to maximize the ratio of variance between two different classes, i.e., left hand and right hand imaginations \citep{Wu2022}. CSP is motivated by the neuroscience discovery that motor activities, both actual and imagined, modulate the $\mu$-rhythm \citep{Blankertz2008}, and can trigger noticeable pattern changes in EEG from different brain regions.

\item Classification. Representative neural networks for EEG signal classification, e.g., EEGNet \citep{Lawhern2018EEGNet}, EEGWaveNet \citep{Thuwajit2022}, and SSVEPNet \citep{Pan2022}, are motivated by our understanding of the corresponding BCI paradigm. For example, EEGNet approximates CSP in designing its convolution layers.
\end{enumerate}

\subsection{EEG Data Augmentation}

Existing EEG data augmentation approaches mainly consider time, frequency, and/or spatial domain transformations.

For time domain transformation, \citet{Wang2018} added random Gaussian white noise to the original signals; \citet{Mohsenvand2020} set a random portion of the EEG signal to zero; and, \citet{Rommel2021} randomly flipped the signals or reversed the axis of time of all channels.

For frequency domain transformation, \citet{Schwabedal2018} randomized the phases of Fourier transforms of all channels; \citet{Mohsenvand2020} and \citet{Cheng2020} randomly filtered a narrow frequency band of all channels; and, \citet{Rommel2021} randomly shifted all channels' power spectral density by a small value.

For spatial domain transformation, \citet{Saeed2021} set the values of some randomly picked channels to zero, or performed random permutation, and \citet{Krell2017} interpolated channels on randomly rotated positions.

Deep learning techniques, e.g., generative adversarial networks \citep{Luo2018}, have also been used for EEG data augmentation \citep{Lashgari2020}.

However, existing EEG data augmentation approaches integrated little task-related knowledge into the data transformation process, which is the challenge to be solved by this paper. The data augmentation approach most similar to ours was \citet{Deiss2018}, which exchanged the left and right hemisphere channels. Section~\ref{sect:algorithms} gives detailed comparisons.

\section{Channel Reflection for EEG Data Augmentation} \label{sect:approach}

Assume the training data include $m$ labeled EEG trials $\{(X^i, y^i)\}_{i=1}^{m}$, where $X^i \in \mathbb{R}^{C \times T}$ is the $i$-th EEG trial ($C$ is the number of EEG channels, and $T$ the number of time samples), and $y^i \in \{0, 1\}$ the corresponding label. Let $X^i_c \in \mathbb{R}^{T}$ be the $c$-th channel of $X^i$, where $c\in\mathbb{N}_{C} =\{1, 2, ..., C\}$. Then, $X^i$ can also be denoted as $X^i = [X^i_1, X^i_2, ..., X^i_{C}]$.

For simplicity, we assume exact symmetric placement of electrodes, i.e., $K$ channels with indices $L=\{L_1,...,L_K\} \subset \mathbb{N}_{C}$ are placed on the left hemisphere, and $K$ channels with indices $R=\{R_1,...,R_K\} \subset \mathbb{N}_{C}$ on the right hemisphere, where $L_k$ and $R_k$ ($k=1,...,K$) are symmetrical electrodes on the left and right hemisphere, respectively.

CR constructs new training trials by exchanging the symmetrical left and right hemisphere channels. More specifically, the left and right hemisphere channels are exchanged, while the middle line channels stay fixed, i.e., the $c$-th channel of the transformed trial $\tilde{X}^i$ becomes
\begin{align}
\tilde{X}^i_c = \begin{cases}
      X^i_{R_k}, & c=L_k \\
      X^i_{L_k}, & c=R_k \\
      X^i_{c}, & c \notin L \cup R \\
    \end{cases}. \label{eq:CR}
\end{align}

The label of $\tilde{X}^i$ is modified in MI classification, but stay identical to the label of $X^i$ for other BCI paradigms, i.e.,
\begin{align}
\tilde{y}^i = \begin{cases}
     1 - y^i, & $left/right hand MI classification$ \\
     y^i, & $ other BCI paradigms $ \\
    \end{cases} \label{eq:CR-label}
\end{align}

\begin{figure*}[htpb]\centering
\subfigure[]{\includegraphics[width=.49\linewidth,clip]{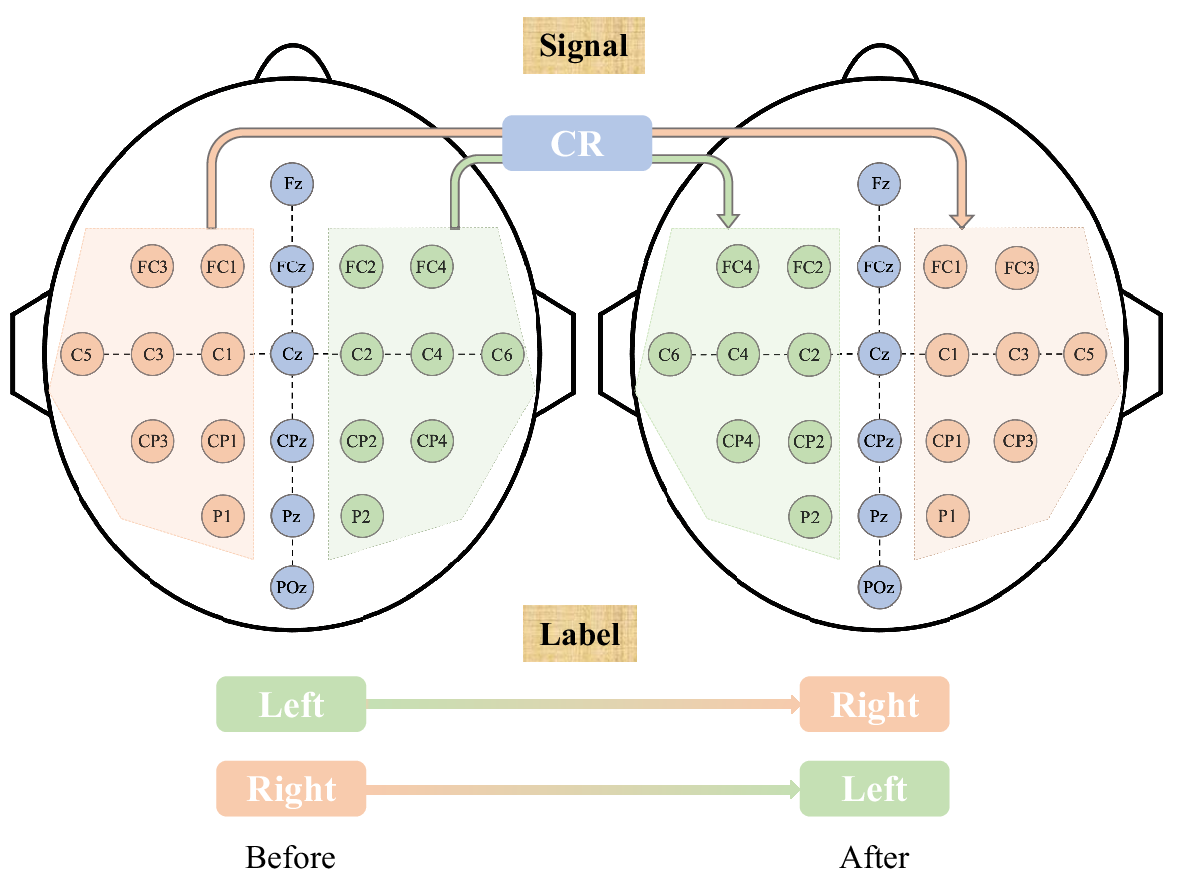}}
\subfigure[]{\includegraphics[width=.49\linewidth,clip]{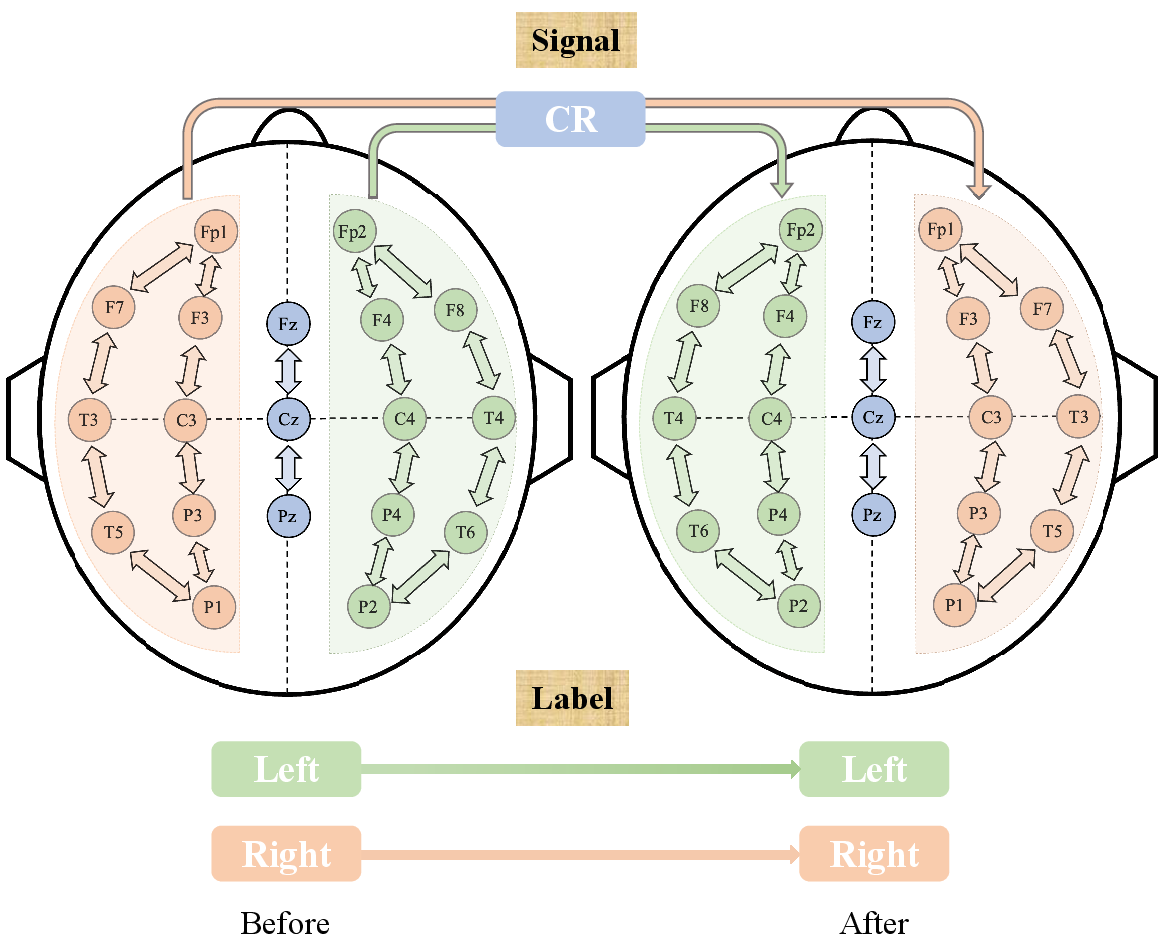}}
\caption{CR data augmentation for (a) unipolar channels, using MI-I dataset \citep{Tangermann2012BNCI2014001} as an example; and, (b) bipolar channels, using Seizure-I dataset \citep{Stevenson2019} as an example.} \label{fig:CR}
\end{figure*}

Figure~\ref{fig:CR} illustrates the details of CR augmentation in MI, SSVEP and P300 paradigms with unipolar channels, and seizure classification with bipolar channels. A unipolar channel [Figure~\ref{fig:CR}(a)] measures the potential difference between an electrode and a common reference, whereas a bipolar channel [Figure~\ref{fig:CR}(b)] outputs the potential difference between two adjacent channels \citep{yao2019reference}. CR augmentation flips all unipolar or bipolar channels from left to right, and right to left.

\begin{figure}[htpb] \centering
\includegraphics[width=\linewidth,clip]{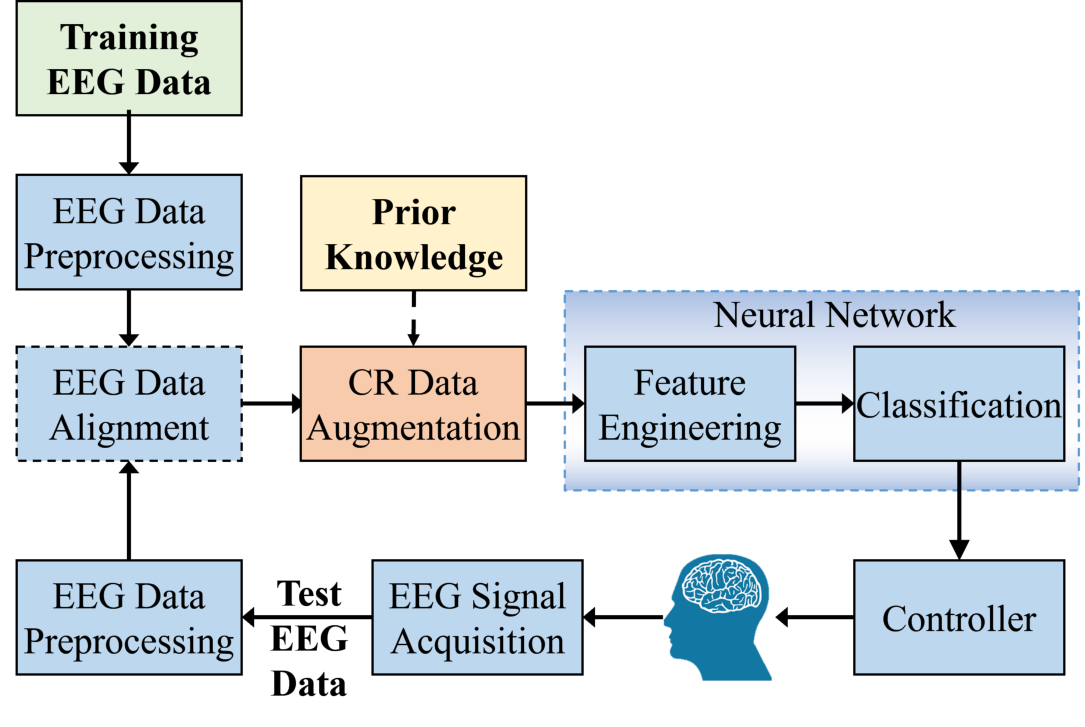}
\caption{Closed-loop EEG-based BCI system, including the proposed CR data augmentation block.} \label{fig:CR_pipeline}
\end{figure}

Figure~\ref{fig:CR_pipeline} shows the flowchart of using CR in a closed-loop EEG-based BCI system, by extending the cross-subject transfer learning pipeline proposed in \citet{Wu2022}. After CR, the training data become the combination of the original EEG trials and the augmented EEG trials.

\section{Experiments and Results}\label{sect:experiments}

Extensive experiments were performed to validate the superior performance of CR. This section presents the experiment settings, results and analyses.

\subsection{Datasets}\label{sect:dataset}

Eight datasets from four different BCI paradigms were used to verify the effectiveness of CR:
\begin{enumerate}
\item MI: Three datasets, namely MI-I \citep{Tangermann2012BNCI2014001} and MI-II \citep{Leeb2007BNCI2014004} from MOABB \citep{Jayaram2018}, and MI-III from BCI Competition IV-1 \citep{Blankertz2007}, were used.
\item SSVEP: The dataset in \citet{nakanishi2015ssvep} was used, and is simply referred to as the SSVEP dataset.
\item P300: Two EEG-based P300 datasets, P300-I \citep{Riccio2013} and P300-II \citep{arico2014erp2}, were used.
\item Seizure classification: Seizure-I \citep{Stevenson2019} and Seizure-II \citep{Wang2023} datasets were used.
\end{enumerate}
Their characteristics are summarized in Table~\ref{tab:datasets}. Table~\ref{tab:LR_channels} also shows the channel locations for each dataset.

\begin{table*}[htpb]  \centering \setlength{\tabcolsep}{1mm}
\caption{Summary of the eight EEG datasets from MI, SSVEP, P300, and seizure classification paradigms.}  \label{tab:datasets}
\begin{tabular}{c|c|c|c|c|c|c}
\toprule
\multirow{2}{*}{Dataset} & Number of & Number of & Sampling & Trial Length & Number of & \multirow{2}{*}{Task Types} \\
 & Subjects & EEG Channels & Rate (Hz) & (seconds) & Total Trials & \\
\midrule
MI-I & 9 & 22 & 250 & 4 & 1,296 & left/right hand \\
MI-II & 9 & 3 & 250 & 4.5 & 1,080 & left/right hand \\
MI-III & 7 & 59 & 250 & 3 & 1,400 & left/right hand, or feet/left hand\\
SSVEP & 10 & 8 & 256 & 1 & 1,800 & 12 different class \\
P300-I & 8 & 8 & 256 & 1 & 29,400 & target/non-target \\
P300-II & 10 & 16 & 256 & 0.8 & 17,280 & target/non-target \\
Seizure-I & 39 & 18 & 256 & 4 & 52,534 & seizure/normal \\
Seizure-II & 27 & 18 & 500 or 1000 & 4 & 21,237 & seizure/normal \\
\bottomrule
\end{tabular}
\end{table*}

\begin{table*}[htbp] \centering \setlength{\tabcolsep}{3mm}
  \caption{Channel locations of the eight datasets.}
    \begin{tabular}{c|l|l}  \toprule
    Dataset & Left Hemisphere & Right Hemisphere \\  \midrule
MI-I & [FC3, FC1, C5, C3, C1, CP3, CP1, P1] & [FC4, FC2, C6, C4, C2, CP4, CP2, P2]\\
\midrule
MI-II & [C3] & [C4] \\
\midrule
\multirow{4}{*}{MI-III}
& [AF3, F5, F3, F1, FC5, FC3, FC1, CFC7, & [AF4, F6, F4, F2, FC6, FC4, FC2, CFC8, \\
& CFC5, CFC3, CFC1, T7, C5, C3, C1, & CFC6, CFC4, CFC2, T8, C6, C4, C2, \\
& CCP7, CCP5, CCP3, CCP1, CP5, CP3, & CCP8, CCP6, CCP4, CCP2, CP6, CP4, \\
& CP1, P5, P3, P1, PO1, O1] & CP2, P6, P4, P2, PO2, O2]\\
\midrule
SSVEP & [P7, P3, O1] & [P8, P4, O2] \\
\midrule
P300-I & [P3, PO7] & [P4, PO8] \\
\midrule
P300-II & [F3, C3, CP3, P3, PO7] & [F4, C4, CP4, P4, PO8]\\
\midrule
\multirow{2}{*}{Seizure-I} & [Fp1-F3, F3-C3, C3-P3, P3-O1, Fp1-F7, & [Fp2-F4, F4-C4, C4-P4, P4-O2, Fp2-F8, \\
& F7-T3, T3-T5, T5-O1] & F8-T4, T4-T6, T6-O2] \\
\midrule
\multirow{2}{*}{Seizure-II}
& [Fp1-F3, F3-C3, C3-P3, P3-O1, Fp1-F7, & [Fp2-F4, F4-C4, C4-P4, P4-O2, Fp2-F8, \\
& F7-T3, T3-T5, T5-O1] & F8-T4, T4-T6, T6-O2] \\
    \bottomrule
    \end{tabular}
  \label{tab:LR_channels}
 \end{table*}

The following data preprocessing procedures were used:
\begin{enumerate}
    \item MI-I and MI-II: The standard preprocessing steps in MOABB, including notch filtering, band-pass filtering, etc., were followed to ensure the reproducibility. For MI-I, only two classes (left/right hand MIs) were used. For both datasets, only EEG trials from the first session of each subject were used for training and test.
    \item MI-III: The EEG data were first [8, 30] Hz band-pass filtered, and then downsampled to 250 Hz to match the other two datasets. Note that Subjects S0 and S5 in MI-III conducted feet/left hand tasks instead of left/right hand tasks; so, their training data were not CR augmented. Again, only EEG trials from the first session of each subject were used for training and test.
    \item SSVEP: We followed the preprocessing steps in \citet{Pan2022}. All signals were first down-sampled to 256 Hz and [6, 80] Hz band-pass filtered via fourth-order forward-backward Butterworth band-pass filter, and then split into 1-second length trials.
    \item P300-I and P300-II: We followed the standard preprocessing steps in MOABB, including band-pass, high-pass and low-pass filtering, etc. Only the first session of P300-I was used, whereas all three sessions of P300-II were used, due to its smaller size.
    \item Seizure-I and Seizure-II: We followed the preprocessing steps in \citep{Wang2023}. Each bipolar EEG channel was preprocessed by a 50 Hz notch filter and a [0.5,50] Hz band-pass filter. EEG signals of Seizure-II were further downsampled to 500 Hz. The signals were then segmented into 4-second long non-overlapping trials.
\end{enumerate}

\subsection{Algorithms} \label{sect:algorithms}

CR was compared with five popular data augmentation baselines. Four of them were described in \citet{Freer2020},  which were also used in \citet{Zhang2022MSDT}. The remaining one was proposed in \citet{Deiss2018}.
\begin{enumerate}
\item No augmentation (Baseline), which does not use any data augmentation.
\item Noise adding (Noise), which adds uniform noise to an EEG trial.
\item Data flipping (Flip), which flips the amplitude of an EEG trial.
\item Data multiplication (Scale), which scales the amplitude of an EEG trial by a coefficient close to 1.
\item Frequency shift (Freq), which uses Hilbert transform \citep{Freeman2007} to shift the frequency of an EEG trial.
\item Channel symmetry (Symm), which reflects left and right hemisphere channels.
\end{enumerate}

Symm seems similar to our proposed CR, but there are two differences:
\begin{enumerate}
\item Symm does not alter the labels, which is crucial for prior knowledge guided data augmentations. For paradigms that are relying detecting event-related desynchronization/synchronization in the left/right hemisphere, the operations of Symm and CR are different.
\item Symm does not explicitly require the left and right hemisphere channels to be strictly symmetric, whereas CR does.
\end{enumerate}

Table~\ref{tab:data_aug} shows the details of the seven approaches. $C_{\text{noise}}=2$, $C_{\text{Scale}}=0.05$ and $C_{\text{freq}}=0.2$ were used, following \citet{Freer2020}. Note that CR does not require any hyperparameters.

\begin{table*}[htbp] \centering
  \caption{Comparison of different data augmentation strategies. }
    \begin{tabular}{c|c|c|c}  \toprule
    Strategy & Formulation & Label & Hyper-parameters \\  \midrule
    Noise  & $\tilde{X}_{c} = X_{c} + rand \ast std(X_{c}) / C_{\text{noise}}$ & Fixed & $C_{\text{noise}}=2$\\
    Flip & $\tilde{X}_{c} = \max(X_{c}) - X_{c}$ & Fixed & --\\
    Scale & $\tilde{X}_{c} = X_{c} \ast (1 \pm C_{\text{scale}})$ & Fixed & $C_{\text{scale}}=0.05$\\
    Freq & $\tilde{X}_{c} = F_{\text{shift}}(X_{c}, \pm C_{\text{freq}})$ & Fixed & $C_{\text{freq}}=0.2$ \\
    Symm & $\tilde{X}_{c} = \begin{cases}
      X^i_{c'}, & c \in L, c' \in R \\
      X^i_{c'}, & c \in R, c' \in L \\
      X^i_{c}, & c \notin \{L \cup R\} \\
    \end{cases}$ & Fixed & --\\
    CR (ours) & $\tilde{X}_{c} = \begin{cases}
      X^i_{R_k}, & c=L_k \\
      X^i_{L_k}, & c=R_k \\
      X^i_{c}, & c \notin \{L \cup R\} \\
    \end{cases}$ & Exchanged (based on task) & --\\
    \bottomrule
    \end{tabular}
  \label{tab:data_aug}
 \end{table*}

\subsection{Experiment Settings}

EEG signals usually exhibit large inter-subject variations (individual differences), and are non-stationary. As a result, collecting labeled calibration data from the target user is generally required for satisfactory classification accuracy \citep{Wu2022}. Three experiment settings with different amounts of training data were used, as illustrated in Figure~\ref{fig:scenario}:

\begin{figure}[htpb] \centering
\includegraphics[width=\linewidth,clip]{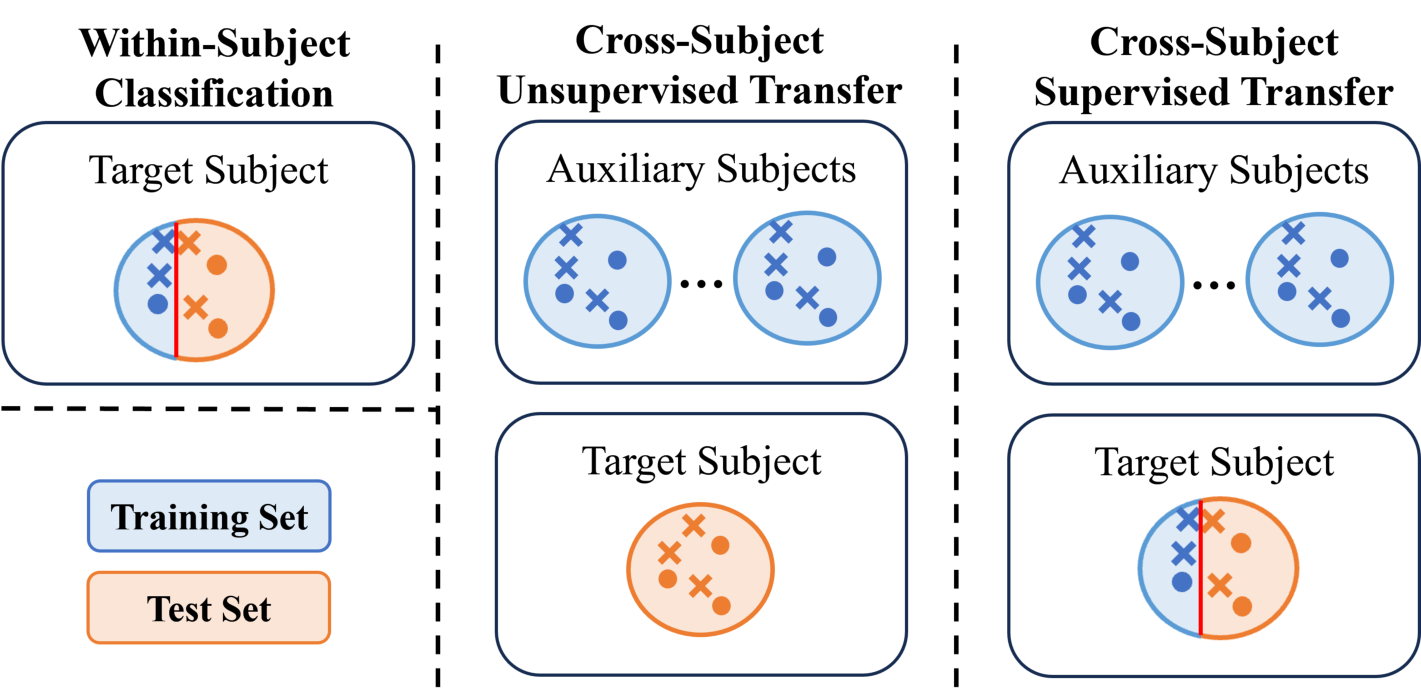}
\caption{Illustration of the three experiment scenarios.} \label{fig:scenario}
\end{figure}

\begin{enumerate}
\item Within-subject classification, where the training set contains only a few labeled data from the target subject, but no data from the source subjects were used.
\item Cross-subject unsupervised transfer, where the training set contains a decent amount of labeled data from the source subjects, but no labeled data from the target subject.
\item Cross-subject supervised transfer, where the training set contains a large amount of labeled data from the source subjects, and a small amount of labeled data from the target subject.
\end{enumerate}

In all three scenarios, only the training set was used to tune the algorithms, and the test set was inaccessible during the training phase. The labels of the test set were used only during the test phase to compute the performance measures. For each dataset, each subject was treated as the target subject once, and all remaining subjects as the source subjects. For the target subject, different amounts of labeled trials in a continuous block [as in \citep{Li2021}] were used in training, with the remaining trials in testing.

In all three scenarios, data augmentation was applied to all labeled trials in the training set to generate extra training data. The amount of training data was doubled for Noise, Flip, Symm and CR, and tripled for Scale and Freq.

The raw classification accuracy was used as the performance measure for MI and SSVEP classification, which usually do not have significant class-imbalance. However, significant class-imbalance exists in the other two paradigms, i.e., the non-target class overwhelmed the target class in P300, and the normal class overwhelmed the seizure class in seizure classification; thus, the raw classification accuracy may be misleading. So, balanced classification accuracy (BCA), defined as the average of recall obtained on each class, was used as the performance measure for P300 and seizure classification.

All experiments using neural networks were repeated five times to accommodate randomness.

\subsection{Main Results}

\subsubsection{MI Classification} \label{sect:miexp}

For the MI paradigm, the classic CSP \citep{Blankertz2008} filtering with Linear Discriminant Analysis (LDA) classifier was used in within-subject classification. CSP used ten spatial filters. In cross-subject unsupervised and supervised transfers, EEGNet \citep{Lawhern2018EEGNet}, a compact convolutional neural network (CNN) with four convolutional layers, was used. In both within-subject classification and cross-subject supervised transfer, the number of labeled trials per class from the target subject increased from 5 to 45 with step 5. The batch size was 32 for Baseline, doubled for Noise, Flip and CR, and tripled for Scale and Freq. All models were trained with 100 epochs using Adam optimizer with learning rate $10^{-3}$, identical to those in \citet{Li2024T-TIME}. The final update steps of gradient descent for each evaluation instance stayed the same.

Euclidean Alignment (EA) \citep{He2020EA}, an effective unsupervised EEG data alignment approach for MI \citep{Wu2022}, was applied before all approaches to align the EEG trials for each source subject. For the target subject, the reference matrix of EA was calculated on the labeled trials, then incrementally updated as each new test trial arrived, as in \citep{Li2024T-TIME}.

Tables~\ref{tab:BNCI2014001}-\ref{tab:MI1} show the results:
\begin{enumerate}
\item Comparing the Baseline performance in the three settings, we can find that when the labeled target data were small, cross-subject unsupervised transfer generally outperformed or performed comparably with within-subject transfer, and cross-subject supervised transfer always outperformed the former two, suggesting the benefits of using source data in transfer learning.
\item The five existing data augmentation approaches did not noticeably improve over Baseline; particularly, Symm significantly degraded the performance, because switching the left/right channels without switching the label contradicts the neuroscience principle of MI.

\item In terms of average performance, our proposed CR always outperformed all other approaches in all three classification scenarios, indicating the effectiveness and robustness of incorporating prior knowledge into data augmentation.
\end{enumerate}

\begin{table*}[h!]     \centering \setlength{\tabcolsep}{1.5mm} \renewcommand\arraystretch{0.86}
       \caption{Classification accuracies  (\%) on MI-I using different data augmentation approaches. The best average performance in each panel is marked in bold, and the second best by an underline.}  \label{tab:BNCI2014001}
    \begin{tabular}{cccccccccccc}   \toprule
Scenario & Approach & S0 & S1 & S2 & S3 & S4 & S5 & S6 & S7 & S8 & Avg. \\
\bottomrule
         & Baseline & 70.44 & 59.33 & 81.33 & 64.11 & 48.78 & 66.89 & 60.78 & 89.11 & 71.11 & 67.99 \\
        & Noise & 69.89 & 58.56 & 81.11 & 63.78 & 49.22 & 67.44 & 62.56 & 89.89 & 70.22 & 68.07 \\
       Within & Flip & 71.78 & 57.89 & 80.89 & 64.67 & 52.33 & 66.44 & 61.78 & 88.78 & 71.44 & \underline{68.44} \\
       -subject & Scale & 70.00 & 58.56 & 82.22 & 62.33 & 49.78 & 66.44 & 61.33 & 89.33 & 70.67 & 67.85 \\
       classification & Freq & 70.00 & 59.44 & 81.22 & 63.44 & 50.78 & 67.56 & 62.00 & 90.33 & 71.22 & \underline{68.44} \\
        & Symm & 52.00 & 56.11 & 52.00 & 58.11 & 52.11 & 51.22 & 57.22 & 68.78 & 63.78 & 56.81 \\
        & CR & 75.67 & 56.00 & 88.22 & 69.78 & 46.78 & 71.22 & 71.00 & 87.11 & 73.67 & \textbf{71.05} \\
        \hline
         & Baseline  & 82.22 & 60.69 & 93.19 & 68.06 & 56.11 & 72.64 & 65.00 & 85.42 & 80.14 & 73.72$_{\pm1.14}$ \\
      Cross  & Noise & 79.17 & 62.08 & 95.69 & 67.08 & 60.42 & 70.28 & 65.14 & 84.17 & 81.81 & 73.98$_{\pm1.06}$ \\
      -subject  & Flip & 75.69 & 59.58 & 94.72 & 65.83 & 57.36 & 75.83 & 62.22 & 83.61 & 82.78 & 73.07$_{\pm0.84}$ \\
      unsupervised  & Scale & 80.56 & 61.11 & 92.78 & 65.56 & 56.94 & 71.25 & 63.75 & 85.42 & 81.53 & 73.21$_{\pm0.76}$ \\
     Transfer   & Freq & 83.47 & 62.22 & 93.61 & 68.33 & 58.75 & 72.22 & 62.92 & 85.83 & 82.08 & \underline{74.38}$_{\pm0.89}$ \\
        & Symm & 52.78 & 52.64 & 54.86 & 54.17 & 51.53 & 52.22 & 52.64 & 55.00 & 53.75 & 53.29$_{\pm0.45}$ \\
        & CR & 84.03 & 63.06 & 91.39 & 71.67 & 60.83 & 75.00 & 70.00 & 85.00 & 80.97 & \textbf{75.77}$_{\pm1.58}$ \\
        \hline
         & Baseline  & 84.96 & 60.93 & 94.53 & 72.51 & 63.60 & 74.11 & 68.01 & 86.94 & 89.39 & 77.22$_{\pm0.81}$ \\
     Cross   & Noise & 83.85 & 59.25 & 93.59 & 72.76 & 63.34 & 72.50 & 68.43 & 86.20 & 88.52 & 76.49$_{\pm1.37}$ \\
     -subject   & Flip & 85.40 & 60.01 & 94.05 & 71.27 & 59.12 & 73.40 & 65.06 & 86.15 & 90.89 & 76.15$_{\pm1.00}$ \\
      supervised  & Scale & 83.04 & 59.54 & 92.40 & 71.49 & 67.81 & 73.73 & 69.06 & 84.77 & 89.23 & 76.78$_{\pm1.46}$ \\
      transfer & Freq & 85.12 & 63.18 & 94.26 & 70.88 & 68.68 & 74.31 & 69.80 & 87.29 & 90.65 & \underline{78.24}$_{\pm1.30}$ \\
        & Symm & 57.01 & 52.43 & 69.54 & 52.27 & 51.10 & 53.00 & 52.61 & 73.86 & 59.39 & 57.91$_{\pm1.20}$ \\
        & CR & 91.35 & 64.30 & 95.18 & 75.76 & 70.60 & 76.06 & 78.51 & 87.31 & 85.12 & \textbf{80.46}$_{\pm1.02}$ \\
\bottomrule
    \end{tabular}
\end{table*}

\begin{table*}[h!]     \centering \setlength{\tabcolsep}{1.2mm} \renewcommand\arraystretch{0.86}
       \caption{Classification accuracies  (\%) on MI-II using different data augmentation approaches. The best average performance in each panel is marked in bold, and the second best by an underline.}  \label{tab:BNCI2014004}
    \begin{tabular}{cccccccccccc}   \toprule
Scenario & Approach & S0 & S1 & S2 & S3 & S4 & S5 & S6 & S7 & S8 & Avg. \\
\bottomrule
         & Baseline  & 82.56 & 59.78 & 48.56 & 77.56 & 58.78 & 70.11 & 61.78 & 55.11 & 66.78 & 64.56 \\
        & Noise & 82.89 & 60.11 & 49.44 & 77.78 & 59.00 & 70.11 & 61.56 & 55.78 & 66.56 & \underline{64.80} \\
      Within  & Flip & 81.11 & 56.22 & 43.00 & 64.11 & 49.78 & 49.67 & 54.78 & 55.44 & 69.11 & 58.14 \\
      -subject  & Scale & 82.44 & 59.00 & 49.22 & 77.89 & 58.78 & 70.33 & 62.00 & 55.78 & 66.78 & 64.69 \\
      classification  & Freq & 82.56 & 59.78 & 48.56 & 77.56 & 58.78 & 70.11 & 61.78 & 55.11 & 66.78 & 64.56 \\
        & Symm & 63.22 & 56.33 & 40.78 & 46.56 & 49.44 & 51.11 & 52.00 & 52.44 & 48.33 & 51.14 \\
        & CR & 83.44 & 56.00 & 56.00 & 78.89 & 57.89 & 71.89 & 59.67 & 50.78 & 73.33 & \textbf{65.32} \\
        \hline
         & Baseline  & 81.33 & 60.83 & 63.00 & 72.67 & 63.83 & 71.67 & 64.00 & 65.67 & 66.33 & 67.70$_{\pm0.91}$ \\
      Cross  & Noise & 81.67 & 62.00 & 64.17 & 72.00 & 66.33 & 70.83 & 64.33 & 64.67 & 67.50 & 68.17$_{\pm0.77}$ \\
      -subject & Flip & 84.00 & 57.83 & 62.50 & 74.33 & 63.17 & 70.50 & 66.00 & 67.83 & 67.67 & \underline{68.20}$_{\pm0.53}$ \\
      unsupervised  & Scale & 79.00 & 58.83 & 66.33 & 69.83 & 65.33 & 70.17 & 62.67 & 68.17 & 64.83 & 67.24$_{\pm1.32}$ \\
       transfer & Freq & 82.67 & 57.83 & 64.00 & 71.83 & 65.17 & 67.50 & 66.67 & 63.50 & 68.83 & 67.56$_{\pm0.55}$ \\
        & Symm & 49.00 & 48.00 & 49.83 & 49.33 & 48.67 & 46.50 & 50.33 & 49.00 & 47.50 & 48.69$_{\pm2.03}$ \\
        & CR & 84.83 & 59.33 & 66.83 & 73.50 & 64.17 & 73.00 & 65.33 & 64.00 & 71.00 & \textbf{69.11}$_{\pm0.84}$ \\
        \hline
         & Baseline  & 88.73 & 57.11 & 56.44 & 76.41 & 66.27 & 75.89 & 68.25 & 59.04 & 71.76 & 68.88$_{\pm1.16}$ \\
        Cross& Noise & 87.80 & 56.89 & 57.20 & 76.04 & 68.47 & 74.58 & 67.03 & 58.49 & 70.34 & 68.54$_{\pm0.68}$ \\
       -subject & Flip & 88.99 & 57.71 & 54.03 & 80.51 & 65.85 & 75.26 & 73.01 & 57.35 & 73.24 & \underline{69.55}$_{\pm1.01}$ \\
        supervised & Scale & 87.46 & 56.55 & 54.93 & 73.32 & 67.11 & 71.80 & 67.33 & 58.22 & 70.74 & 67.49$_{\pm0.87}$ \\
       transfer & Freq & 87.91 & 57.23 & 56.39 & 80.05 & 65.77 & 72.78 & 68.53 & 57.43 & 73.36 & 68.83$_{\pm0.94}$ \\
        & Symm & 53.62 & 48.81 & 52.52 & 57.20 & 48.18 & 46.50 & 53.16 & 52.11 & 46.90 & 51.00$_{\pm2.31}$ \\
        & CR & 90.13 & 62.35 & 59.22 & 76.56 & 64.80 & 76.95 & 70.89 & 57.62 & 71.65 & \textbf{70.02}$_{\pm1.13}$ \\
\bottomrule
    \end{tabular}
\end{table*}

\begin{table*}[htpb]     \centering \setlength{\tabcolsep}{2mm} \renewcommand\arraystretch{0.9}
       \caption{Classification accuracies  (\%) on MI-III using different data augmentation approaches. The best average performance in each panel is marked in bold, and the second best by an underline.}  \label{tab:MI1}
    \begin{tabular}{cccccccccc}   \toprule
Scenario & Approach & S0 & S1 & S2 & S3 & S4 & S5 & S6 & Avg. \\
\bottomrule
        & Baseline  & 66.89 & 57.33 & 62.89 & 81.67 & 84.22 & 74.89 & 80.78 & 72.67 \\
        & Noise & 68.00 & 55.56 & 61.11 & 82.00 & 84.56 & 74.67 & 78.67 & 72.08 \\
     Within   & Flip & 69.22 & 54.67 & 61.44 & 83.78 & 87.89 & 73.78 & 77.11 & 72.56 \\
     -subject   & Scale & 67.22 & 55.56 & 62.89 & 83.44 & 84.56 & 75.22 & 81.22 & \underline{72.87} \\
      classification  & Freq & 67.56 & 55.00 & 62.67 & 83.00 & 84.56 & 72.22 & 79.11 & 72.02 \\
        & Symm & 66.89 & 54.78 & 65.11 & 74.67 & 57.78 & 74.89 & 57.67 & 64.54 \\
        & CR & 66.89 & 59.22 & 61.33 & 81.78 & 85.00 & 74.89 & 89.22 & \textbf{74.05} \\
        \hline
          & Baseline  & 66.00 & 69.30 & 65.70 & 58.30 & 92.20 & 74.80 & 71.90 & 71.17$_{\pm0.87}$ \\
      Cross  & Noise & 69.60 & 69.70 & 65.80 & 59.40 & 92.80 & 75.20 & 70.60 & \underline{71.87}$_{\pm0.55}$ \\
      -subject  & Flip & 67.50 & 66.70 & 66.10 & 54.50 & 91.50 & 75.00 & 67.70 & 69.86$_{\pm1.49}$ \\
       unsupervised & Scale & 68.80 & 69.40 & 66.10 & 60.50 & 91.20 & 73.30 & 72.60 & 71.70$_{\pm0.73}$ \\
       transfer & Freq & 68.10 & 68.30 & 66.00 & 58.90 & 91.30 & 73.70 & 70.40 & 70.96$_{\pm0.82}$ \\
        & Symm & 53.50 & 49.00 & 49.80 & 50.60 & 47.40 & 53.80 & 42.50 & 49.51$_{\pm0.80}$ \\
        & CR & 66.40 & 75.60 & 64.50 & 68.10 & 92.90 & 72.10 & 85.20 & \textbf{74.97}$_{\pm0.96}$ \\
        \hline
        & Baseline  & 75.16 & 74.34 & 69.98 & 75.57 & 93.48 & 81.51 & 80.13 & \underline{78.60}$_{\pm1.13}$ \\
     Cross   & Noise & 76.26 & 72.70 & 71.37 & 74.86 & 92.85 & 78.85 & 79.69 & 78.08$_{\pm1.00}$ \\
     -subject   & Flip & 73.55 & 73.41 & 70.55 & 73.99 & 94.16 & 77.69 & 77.11 & 77.21$_{\pm1.66}$ \\
     supervised   & Scale & 76.16 & 71.11 & 68.61 & 70.50 & 91.03 & 77.56 & 75.86 & 75.83$_{\pm1.04}$ \\
      transfer  & Freq & 75.54 & 72.42 & 69.62 & 71.92 & 90.86 & 78.02 & 77.36 & 76.53$_{\pm0.86}$ \\
        & Symm & 56.88 & 50.33 & 49.33 & 59.29 & 75.28 & 60.66 & 64.31 & 59.44$_{\pm1.78}$ \\
        & CR & 76.86 & 77.65 & 67.31 & 82.11 & 94.32 & 79.14 & 85.91 & \textbf{80.47}$_{\pm0.82}$ \\
\bottomrule
    \end{tabular}
\end{table*}

\begin{table*}[htpb]     \centering \setlength{\tabcolsep}{1mm} \renewcommand\arraystretch{0.95}
       \caption{Classification accuracies (\%) on SSVEP using different data augmentation approaches. The best average performance in each panel is marked in bold, and the second best by an underline.}  \label{tab:ssvep}
    \begin{tabular}{ccccccccccccc}   \toprule
Scenario & Approach & S0 & S1 & S2 & S3 & S4 & S5 & S6 & S7 & S8 & S9 & Avg.\\
\bottomrule
    & Baseline & 79.88 & 60.75 & 80.46 & 93.90 & 92.66 & 94.06 & 92.91 & 97.50 & 92.32 & 74.42 & 85.89$_{\pm0.15}$ \\
        Within& Noise & 65.95 & 47.29 & 69.85 & 90.34 & 88.6 & 91.38 & 92.16 & 97.71 & 88.37 & 68.1 & 79.97$_{\pm0.50}$  \\
      -subject  & Flip & 66.48 & 47.87 & 69.5 & 90.48 & 88.77 & 91.68 & 91.98 & 97.86 & 89.06 & 68.26 & 80.19$_{\pm0.39}$  \\
      classification  & Scale & 81.57 & 62.44 & 83.6 & 93.22 & 94.28 & 94.92 & 93.06 & 97.32 & 93.28 & 75.36 & \textbf{86.91}$_{\pm0.07}$  \\
        & Freq & 71.55 & 54.49 & 78.38 & 91.51 & 91.66 & 91.86 & 92.59 & 92.56 & 89.87 & 63.91 & 81.84$_{\pm0.30}$ \\
        & CR & 81.56 & 64.07 & 79.95 & 94.57 & 94.25 & 93.83 & 92.86 & 97.22 & 92.19 & 76.26 & \underline{86.68}$_{\pm0.28}$\\
        \hline
        & Baseline & 52.52 & 43.77 & 61.87 & 95.81 & 95.03 & 93.88 & 94.39 & 97.90 & 95.27 & 86.97 & \underline{81.74}$_{\pm0.37}$\\
   Cross     & Noise & 42.78  & 46.52  & 47.37  & 97.21  & 93.32  & 94.16  & 92.21  & 93.75  & 91.24  & 78.20  & 77.68$_{\pm0.20}$   \\
    -subject    & Flip & 32.22  & 34.03  & 42.64  & 89.58  & 89.45  & 89.72  & 85.14  & 95.97  & 87.91  & 74.17  & 72.08$_{\pm0.38}$   \\
    unsupervised    & Scale & 46.67  & 45.37  & 50.74  & 97.59  & 94.63  & 94.26  & 91.30  & 95.19  & 89.07  & 80.56  & 78.54$_{\pm0.46}$   \\
     transfer   & Freq & 45.18  & 38.34  & 49.08  & 95.00  & 93.52  & 93.33  & 92.78  & 94.26  & 81.30  & 81.30  & 76.41$_{\pm0.23}$  \\
        & CR & 56.89 & 48.11 & 67.33 & 95.33 & 96.11 & 91.67 & 94.78 & 98.67 & 93.66 & 88.67 & \textbf{83.12}$_{\pm0.17}$ \\
        \hline
        & Baseline & 67.88 & 54.11 & 79.03 & 97.30 & 97.88 & 95.35 & 93.28 & 99.09 & 95.57 & 81.25 & 86.07$_{\pm0.30}$ \\
  Cross      & Noise & 69.04  & 52.28  & 78.83  & 97.26  & 97.80  & 95.88  & 92.49  & 98.32  & 95.36  & 81.40  & 85.87$_{\pm0.23}$   \\
  -subject      & Flip & 63.66  & 43.83  & 65.64  & 94.52  & 84.68  & 91.74  & 92.64  & 96.16  & 94.85  & 77.05  & 80.48$_{\pm0.27}$   \\
  supervised      & Scale & 72.34  & 56.52  & 83.85  & 97.07  & 97.20  & 96.50  & 94.11  & 98.05  & 95.87  & 85.69  & 87.72$_{\pm0.11}$   \\
  transfer      & Freq & 73.05  & 54.69  & 86.52  & 97.54  & 97.95  & 94.29  & 96.25  & 97.53  & 95.34  & 84.95  & \underline{87.81}$_{\pm0.22}$  \\
        & CR & 74.54 & 58.33 & 84.64 & 97.33 & 97.93 & 93.99 & 95.74 & 98.97 & 94.59 & 87.99 & \textbf{88.40}$_{\pm0.28}$ \\
\bottomrule
    \end{tabular}
\end{table*}

\subsubsection{SSVEP Classification} \label{sect:ssvepexp}

Three different settings were also considered for SSVEP.

SSVEPNet \citep{Pan2022} trained with 500 epochs and Adam optimizer was used as the classifier. The number of labeled target trials per class increased from 1 to 9 with step 1. For baseline, batch size 30 and learning rate $10^{-2}$ were used in within-subject classification, and batch size 64 and learning rate $10^{-3}$ in cross-subject transfers. The batch size was doubled for Noise, Flip and CR, and tripled for Scale and Freq.

Table~\ref{tab:ssvep} shows the results:
\begin{enumerate}
\item Unlike the case in MI classification, here cross-subject unsupervised transfer always had worse performance than within-subject classification. This may be because SSVEP had 12 classes, much more than the two classes in MI, so it was more sensitive to the data discrepancies between the source and target subjects.
\item Although most of the four existing data augmentation approaches were ineffective, our proposed CR always outperformed Baseline, again indicating the effectiveness and robustness of incorporating prior knowledge into data augmentation.
\end{enumerate}

\subsubsection{P300 Classification} \label{sect:p300exp}

Due to page limit, only cross-subject unsupervised transfer was consider in P300. EEGNet with batch size 256 was used. All other hyperparameters were identical to those in MI.

Tables~\ref{tab:p300_1} and \ref{tab:p300_2} show the results. All data augmentation approaches were effective, but our proposed CR achieved the best or second best average BCAs, comparable to Noise. However, Noise has one hyper-parameter, whereas CR is completely parameter-free.

\begin{table*}[htpb]     \centering \setlength{\tabcolsep}{3mm} \renewcommand\arraystretch{1}
       \caption{BCAs (\%) of different data augmentation approaches in cross-subject unsupervised transfer on P300-I. The best average performance is marked in bold, and the second best by an underline.}  \label{tab:p300_1}
    \begin{tabular}{cccccccccc}   \toprule
Approach & S0 & S1 & S2 & S3 & S4 & S5 & S6 & S7& Avg.\\
\bottomrule
Baseline & 71.65 & 69.35 & 75.11 & 70.31 & 73.46 & 69.94 & 74.18 & 72.78  & 72.10$_{\pm0.17}$  \\
Noise & 71.17 & 70.27 & 75.59 & 70.33 & 74.00 & 70.29 & 74.56 & 74.29  & \underline{72.56}$_{\pm0.23}$  \\
Flip & 71.43 & 69.61 & 75.92 & 70.19 & 73.99 & 70.67 & 74.14 & 73.42  & 72.42$_{\pm0.21}$  \\
Scale & 71.09 & 70.36 & 75.45 & 70.34 & 73.89 & 70.08 & 74.13 & 74.88  & 72.53$_{\pm0.21}$  \\
Freq & 71.60 & 70.44 & 74.42 & 69.86 & 74.73 & 69.57 & 74.76 & 73.92  & 72.41$_{\pm0.23}$  \\
CR & 71.50 & 70.35 & 75.43 & 70.21 & 74.36 & 70.66 & 74.82 & 74.33 & \textbf{72.71}$_{\pm0.13}$ \\
\bottomrule
\end{tabular}
\end{table*}

\begin{table*}[t]     \centering \setlength{\tabcolsep}{2mm} \renewcommand\arraystretch{1}
       \caption{BCAs (\%) of different data augmentation approaches in cross-subject unsupervised transfer on P300-II. The best average performance is marked in bold, and the second best by an underline.}  \label{tab:p300_2}
    \begin{tabular}{cccccccccccc}   \toprule
Approach & S0 & S1 & S2 & S3 & S4 & S5 & S6 & S7& S8 & S9 & Avg.\\
\bottomrule
        Baseline & 76.04 & 85.51 & 79.26 & 85.52 & 87.15 & 79.06 & 76.92 & 74.28 & 87.95 & 87.71 & 81.94$_{\pm0.16}$  \\
        Noise & 78.36 & 85.10 & 79.66 & 85.65 & 88.08 & 79.15 & 80.18 & 76.29 & 88.35 & 89.50 & \textbf{83.03}$_{\pm0.08}$  \\
        Flip & 76.70 & 85.26 & 79.11 & 84.54 & 88.78 & 79.86 & 78.65 & 75.54 & 88.73 & 87.54 & 82.47$_{\pm0.20}$  \\
        Scale & 78.52 & 85.13 & 79.76 & 85.32 & 88.09 & 78.63 & 79.38 & 75.25 & 88.26 & 89.14 & 82.75$_{\pm0.28}$  \\
        Freq & 78.45 & 86.01 & 79.63 & 84.91 & 88.17 & 80.05 & 79.50 & 75.70 & 88.73 & 88.69 & 82.98$_{\pm0.12}$  \\
        CR & 77.91 & 85.39 & 79.69 & 85.92 & 88.32 & 79.51 & 79.81 & 75.52 & 88.63 & 89.53 & \underline{83.02}$_{\pm0.11}$ \\
\bottomrule
\end{tabular}
\end{table*}

\subsubsection{Seizure Classification} \label{sect:seizureexp}

EEGNet and EEGWaveNet \citep{Thuwajit2022} were used in cross-subject unsupervised transfer in seizure classification. Same as \citet{Wang2023}, both networks were trained for 100 epochs. Batch size 256 was used for Baseline. It was doubled for Noise, Flip and CR, and tripled for Scale and Freq.

Tables~\ref{tab:seizure_1} and \ref{tab:seizure_2} show the results. Again, our proposed CR achieved the best average BCAs. Particularly, it was the only effective data augmentation approach for EEGNet.
\begin{table*}[htpb]  \centering \setlength{\tabcolsep}{1.5mm} \renewcommand\arraystretch{1}
       \caption{BCAs (\%) of different data augmentation approaches on Seizure-I in cross-subject unsupervised transfer. The best average performance is marked in bold, and the second best by an underline.}  \label{tab:seizure_1}
\begin{tabular}{c|cccccc|cccccc}   \toprule
\multirow{2}{*}{ID}&\multicolumn{6}{c}{EEGNet}&\multicolumn{6}{c}{EEGWaveNet}\\ \cline{2-13}
& Baseline & Noise & Flip & Scale & Freq & CR & Baseline & Noise & Flip & Scale & Freq & CR\\
\bottomrule
S1&50.98&62.07&62.00&56.72&61.38&55.73&59.67&54.21&57.47&58.58&62.53&65.55\\
S4&81.01&60.37&60.55&60.60&57.02&66.03&70.02&67.59&72.85&67.77&70.09&69.52\\
S5&69.26&79.00&78.27&78.92&78.78&77.38&70.39&65.28&73.08&70.45&64.58&74.47\\
S7&57.38&55.82&63.15&60.86&61.16&66.30&70.78&65.68&62.72&63.18&61.99&66.96\\
S9&74.30&55.94&57.12&52.09&55.53&54.94&64.39&65.56&57.35&65.51&67.64&67.62\\
S11&59.65&63.95&70.85&64.32&71.29&70.01&62.03&61.57&61.40&60.50&70.68&58.74\\
S13&60.37&54.75&57.18&59.33&55.51&67.13&65.92&60.46&56.67&64.03&57.67&71.20\\
S14&55.02&56.56&56.36&55.30&49.10&49.31&47.17&54.25&56.85&56.42&55.81&52.98\\
S15&51.21&55.60&53.34&54.32&53.30&50.27&51.04&51.64&53.90&50.50&49.80&52.42\\
S16&51.54&51.37&53.85&54.31&54.57&59.61&56.48&62.91&57.78&61.39&62.02&56.81\\
S17&59.55&57.47&63.54&57.65&60.09&52.10&49.98&53.81&50.89&55.09&55.06&53.58\\
S19&62.16&51.65&52.90&48.49&50.40&56.60&52.44&55.08&56.27&57.05&58.03&61.59\\
S20&50.88&59.80&59.90&55.64&61.65&56.78&60.63&58.75&56.76&60.18&53.87&58.26\\
S21&59.23&56.36&87.48&86.19&88.74&89.00&58.97&74.93&68.00&66.09&66.41&61.67\\
S22&40.38&52.79&52.84&51.82&50.38&54.73&53.14&49.92&52.72&55.29&52.02&57.04\\
S25&64.29&71.02&65.89&69.14&66.20&74.60&62.64&59.46&55.93&65.44&62.36&61.77\\
S31&78.75&55.22&57.21&52.87&53.26&59.20&76.75&60.18&59.14&82.31&67.31&87.11\\
S34&75.29&64.86&69.42&57.60&56.62&70.90&76.68&86.57&81.75&93.03&90.07&91.82\\
S36&69.61&78.63&84.57&63.11&75.09&74.00&77.25&73.63&62.48&80.67&79.84&71.97\\
S38&66.76&61.98&58.02&57.39&51.48&54.34&59.70&58.13&58.28&55.38&56.69&60.80\\
S39&74.43&66.78&65.85&64.49&65.09&58.13&69.54&64.02&67.54&71.56&64.47&64.41\\
S40&51.15&49.44&50.62&50.21&50.27&58.84&62.07&56.56&59.85&53.39&64.14&61.99\\
S41&55.26&63.23&61.85&53.02&57.53&64.96&70.64&61.74&66.09&67.81&67.42&65.22\\
S44&68.79&45.39&64.38&42.51&66.24&64.85&64.57&79.30&52.59&69.43&71.01&61.58\\
S47&83.12&63.45&59.04&53.27&62.49&63.73&70.34&64.71&66.71&82.84&71.09&84.48\\
S50&69.92&65.61&69.58&65.37&68.64&73.96&73.85&78.82&70.84&73.62&64.58&68.48\\
S51&54.03&44.26&47.87&46.81&48.37&46.62&55.35&50.41&39.81&61.28&51.48&56.30\\
S52&47.28&60.41&57.31&67.96&56.17&63.85&54.14&50.46&51.41&59.99&56.79&63.88\\
S62&90.19&51.30&53.17&50.05&52.32&46.00&68.78&83.73&76.66&88.12&93.98&88.09\\
S63&50.10&52.71&51.36&54.65&54.19&51.49&48.33&51.69&48.40&54.33&50.95&53.49\\
S66&63.95&71.12&56.07&70.19&65.85&75.69&64.20&66.64&68.66&63.69&60.71&67.71\\
S67&59.71&67.83&70.78&71.32&65.35&76.54&74.97&72.60&63.95&65.32&58.11&74.49\\
S69&50.05&49.98&50.64&68.48&49.93&50.08&83.02&84.68&86.52&89.66&87.90&90.91\\
S73&50.03&60.97&59.50&51.08&50.67&67.94&51.28&60.33&67.20&57.25&56.51&64.82\\
S75&52.47&51.51&50.35&52.87&52.59&56.34&56.01&52.24&58.61&59.15&56.64&59.04\\
S76&53.95&51.79&51.24&55.00&50.65&53.82&53.37&54.67&53.51&52.77&52.98&59.41\\
S77&50.58&58.29&56.61&58.58&54.84&45.98&48.97&52.80&54.11&49.66&49.74&49.43\\
S78&67.16&67.51&69.64&66.34&63.74&80.51&74.13&66.41&63.15&68.56&74.64&76.40\\
S79&57.67&49.57&46.56&49.96&46.32&60.95&62.12&60.79&57.60&62.25&54.40&63.57\\
\hline
\multirow{2}{*}{Avg.}&\underline{61.22}&58.88&60.43&58.69&58.79&\textbf{62.03}&62.86&62.88&61.17&\underline{65.12}&63.38&\textbf{66.04}\\
&${\pm0.54}$&${\pm0.29}$&${\pm0.24}$&${\pm0.43}$&${\pm0.51}$&${\pm0.40}$&${\pm0.68}$&${\pm0.50}$&${\pm0.47}$&${\pm0.71}$&${\pm0.53}$&${\pm0.16}$\\
\bottomrule
\end{tabular}
\end{table*}

\begin{table*}[htpb]  \centering \setlength{\tabcolsep}{1.5mm} \renewcommand\arraystretch{1}
       \caption{BCAs (\%) of different data augmentation approaches on Seizure-II in cross-subject unsupervised transfer setting. The best average performance is marked in bold, and the second best by an underline.}  \label{tab:seizure_2}
\begin{tabular}{c|cccccc|cccccc}   \toprule
\multirow{2}{*}{ID}&\multicolumn{6}{c}{EEGNet}&\multicolumn{6}{c}{EEGWaveNet}\\ \cline{2-7} \cline{8-13}
& Baseline & Noise & Flip & Scale & Freq & CR & Baseline & Noise & Flip & Scale & Freq & CR\\
\bottomrule
S1&89.41&86.37&84.80&86.28&84.22&89.31&83.18&83.27&87.93&91.00&83.49&85.22\\
S2&84.52&81.78&71.15&82.97&78.71&86.04&76.60&76.28&72.85&88.74&66.53&75.00\\
S3&99.17&96.50&93.87&98.83&98.33&97.50&92.36&98.67&95.33&94.12&96.50&98.25\\
S4&92.47&93.55&94.62&93.01&93.01&94.09&94.62&95.16&89.25&92.47&97.31&93.55\\
S5&85.42&80.16&73.05&81.16&77.67&78.35&70.20&75.68&81.57&75.14&73.96&88.54\\
S6&70.67&70.02&76.76&78.13&67.65&78.24&83.38&86.79&74.41&86.81&76.67&84.14\\
S7&93.75&79.58&89.58&93.75&93.75&79.17&85.83&91.67&87.50&67.92&78.33&85.83\\
S8&84.62&78.21&76.23&85.90&76.92&88.46&92.79&90.22&90.22&87.66&94.07&83.23\\
S9&91.03&88.78&94.87&94.55&93.27&93.59&95.19&95.83&91.76&95.83&87.50&91.03\\
S10&82.52&84.53&62.27&67.91&79.43&81.25&57.74&70.70&61.87&66.39&62.48&63.93\\
S11&86.60&91.18&92.75&94.93&94.55&96.50&90.33&85.18&76.03&86.71&84.23&85.39\\
S12&54.11&53.56&52.00&54.26&50.60&60.50&55.06&52.99&52.96&55.95&60.94&66.31\\
S13&85.89&88.74&88.33&94.30&87.22&83.26&66.67&75.00&78.33&62.78&74.44&81.11\\
S14&89.50&89.75&89.88&84.53&85.59&96.55&78.96&71.79&94.67&71.55&93.63&85.13\\
S15&92.19&91.67&92.83&87.76&88.31&67.81&91.31&70.49&72.04&70.56&70.80&55.75\\
S16&82.83&89.39&87.37&89.90&85.86&85.86&78.28&77.27&81.31&81.82&78.28&82.32\\
S17&94.39&93.26&89.54&92.50&90.67&92.05&86.56&81.00&84.22&84.16&80.94&89.77\\
S18&95.42&82.31&97.14&90.67&94.22&97.86&97.22&99.08&92.86&98.78&97.14&84.14\\
S19&82.67&78.47&71.34&80.63&76.32&91.35&64.01&59.87&76.44&66.05&79.73&81.18\\
S20&69.11&62.12&60.04&59.73&60.12&71.02&88.91&86.74&80.84&84.56&81.35&76.77\\
S21&72.06&91.36&75.07&80.37&84.17&86.14&64.49&57.36&73.30&80.96&65.24&80.03\\
S22&89.82&87.50&81.94&88.89&86.11&85.83&87.96&88.15&91.11&84.44&84.35&88.06\\
S23&57.60&58.82&56.87&59.74&55.66&59.69&49.92&59.33&56.39&59.04&50.27&48.05\\
S24&83.38&78.50&81.85&81.57&81.30&84.91&76.62&78.62&81.13&74.37&80.37&82.39\\
S25&92.48&95.51&94.01&95.51&89.14&88.39&93.74&88.61&89.97&94.16&90.29&97.93\\
S26&46.91&45.13&51.79&49.40&55.29&62.00&56.01&64.30&65.26&71.91&56.17&56.97\\
S27&63.45&60.31&65.43&59.88&63.62&59.00&49.46&51.72&54.22&59.68&51.66&55.66\\
\hline
\multirow{2}{*}{Avg.}&\underline{81.92}&80.63&79.46&81.74&80.43&\textbf{82.77}&78.05&78.21&\underline{79.03}&79.02&77.65&\textbf{79.47}\\
&${\pm0.76}$&${\pm0.81}$&${\pm0.78}$&${\pm0.31}$&${\pm0.93}$&${\pm0.40}$&${\pm0.34}$&${\pm0.97}$&${\pm0.70}$&${\pm1.56}$&${\pm1.31}$&${\pm0.57}$\\
\bottomrule
\end{tabular}
\end{table*}

\subsection{Visualizations}\label{sect:visualizations}

\begin{figure*}[h!]\centering
\subfigure[]{\includegraphics[width=\linewidth,clip]{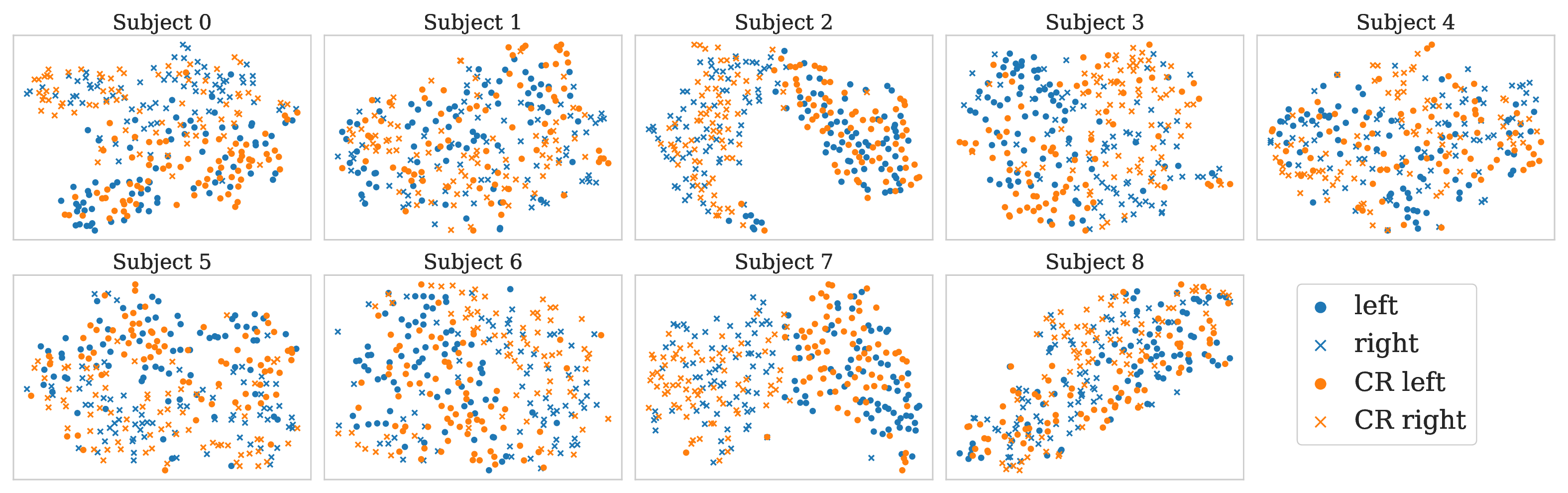}\label{fig:5a}}
\subfigure[]{\includegraphics[width=\linewidth,clip]{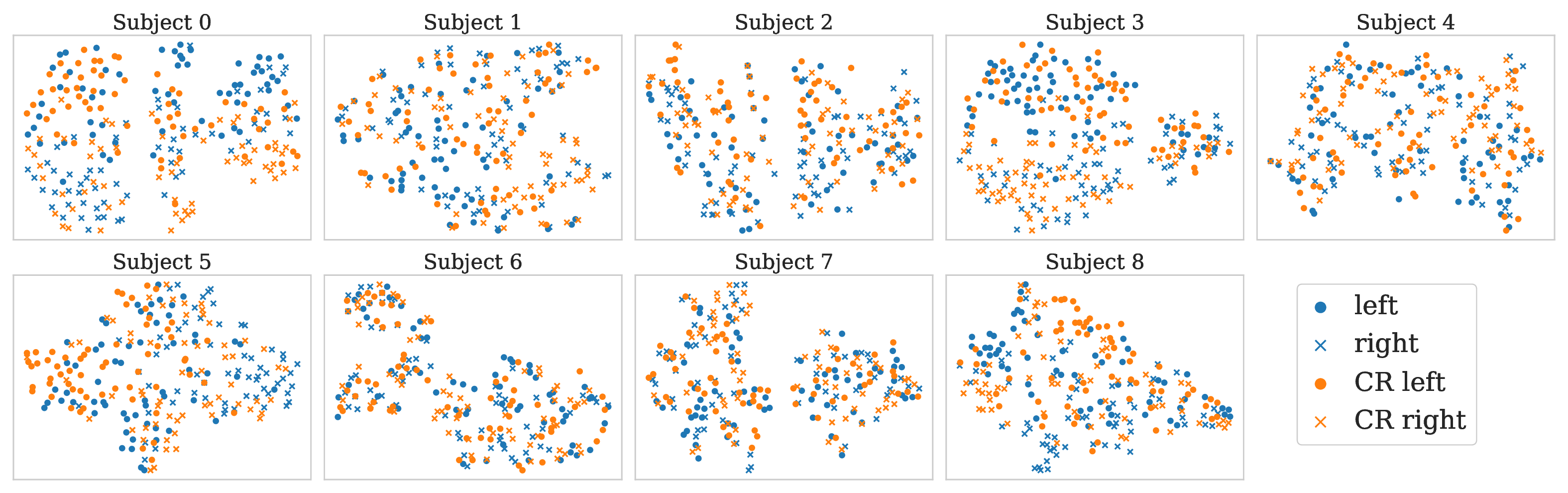}}
\subfigure[]{\includegraphics[width=\linewidth,clip]{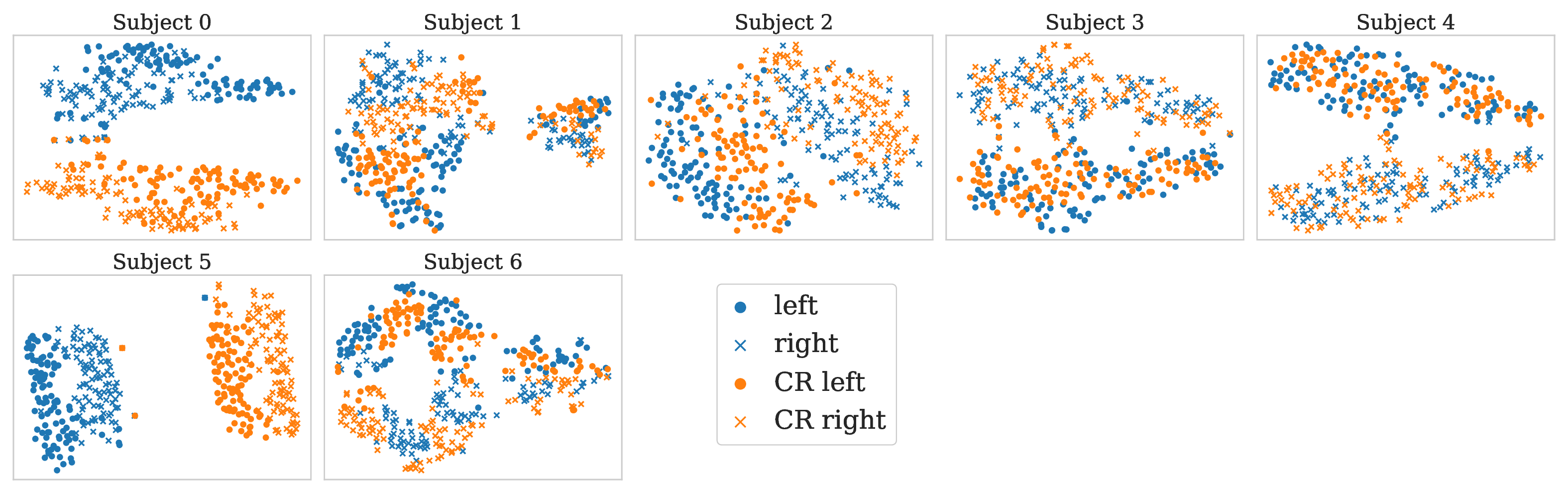}\label{fig:mi1_sub}}
\caption{$t$-SNE visualization of CSP features from the original EEG trials and the CR augmented EEG trials. (a) Subjects S0-S8 in MI-I; (b) Subjects S0-S8 in MI-II; and, (c) Subjects S0-S6 in MI-III. Different colors represent different classes. The dots represent the original trials, and the crosses represent the CR augmented trials. Note that for S0 and S5 in MI-III, the classification task was right hand versus feet.} \label{fig:tsne_subject}
\end{figure*}

Figure~\ref{fig:tsne_subject} uses $t$-SNE \citep{VanderMaaten2008} to visualize the feature distributions of the left and right hand imaginations from some subject in the three MI datasets. Observe that the augmented samples may occur in regions where no original samples were present, which may not be possible without utilizing prior knowledge.

\begin{figure*}[htpb]\centering
\subfigure[]{\includegraphics[width=.9\linewidth,clip]{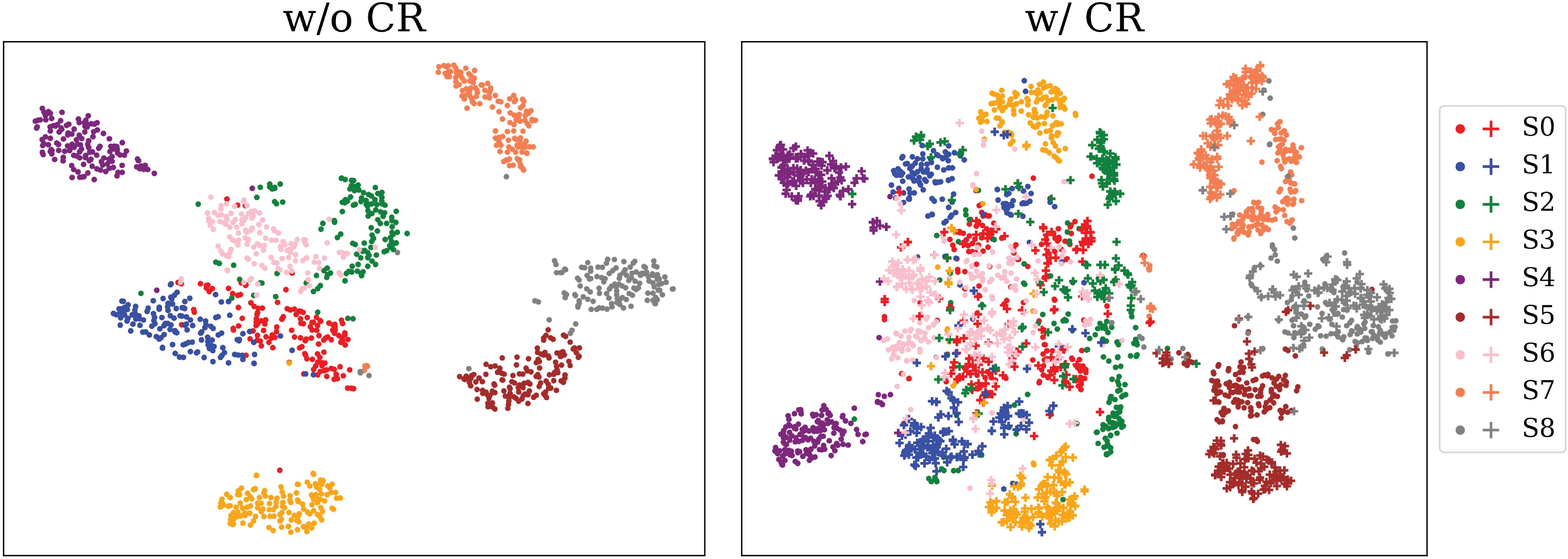} \label{fig:14001_all}}
\subfigure[]{\includegraphics[width=.9\linewidth,clip]{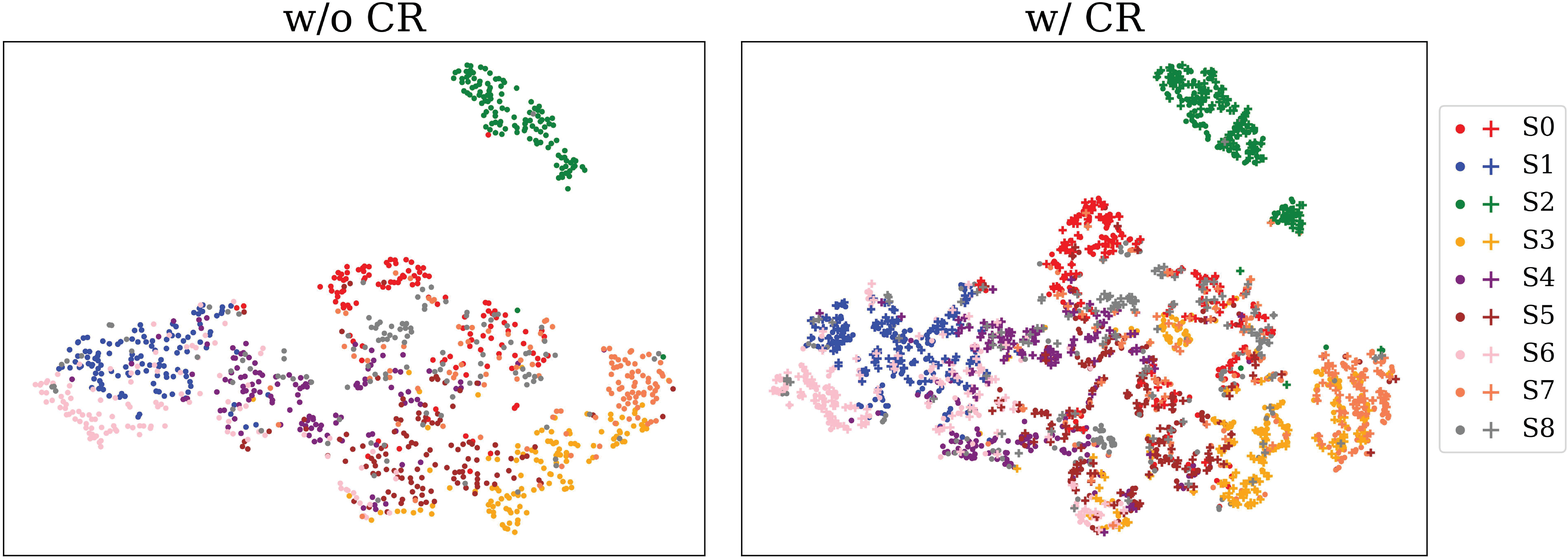}}
\subfigure[]{\includegraphics[width=.9\linewidth,clip]{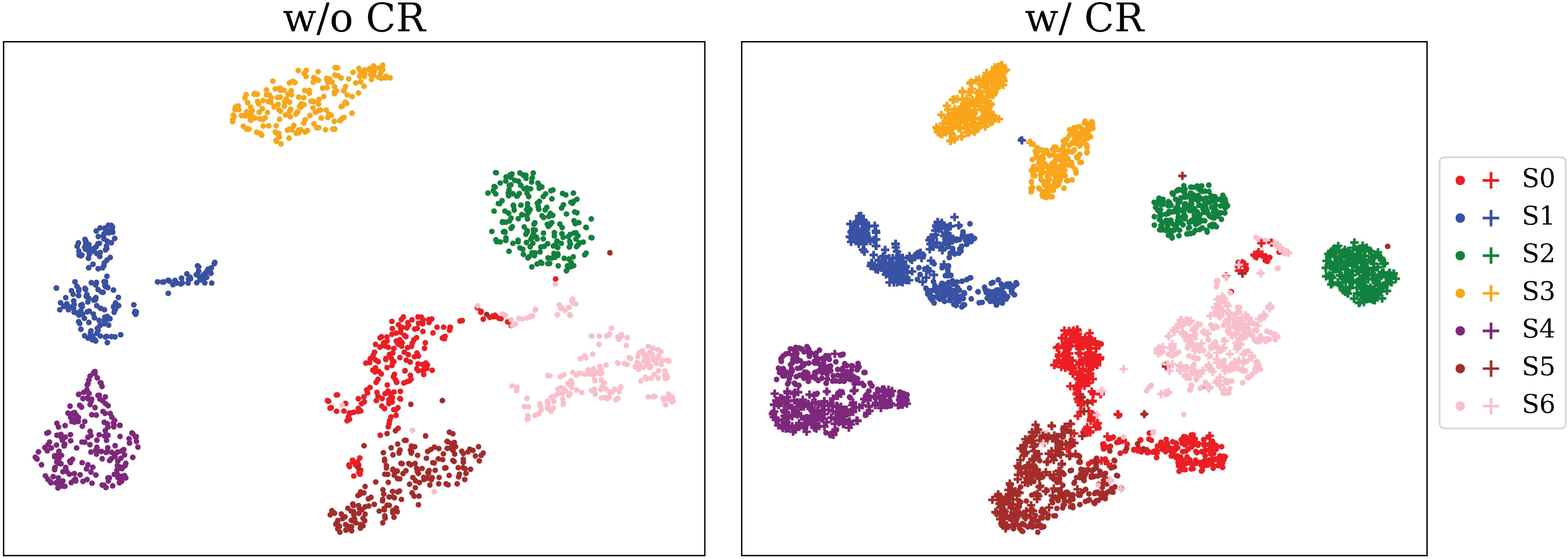}}
\caption{$t$-SNE visualization of CSP features from the original EEG trials and the CR augmented ones. (a) MI-I; (b) MI-II; and, (c) MI-III. Different colors represent different subjects. The dots represent the original trials, and the crosses represent the CR augmented trials.} \label{fig:tsne_all}
\end{figure*}

Figure~\ref{fig:tsne_all} shows $t$-SNE visualization of feature distributions of all subjects on the three MI datasets. Integrating the CR augmented samples added additional information beyond the original samples.

Figures~\ref{fig:tsne_subject} and \ref{fig:tsne_all} show that, for each subject, the CR augmented samples generally had consistent distributions with the original samples. There are a few exceptions:
\begin{enumerate}
\item Figure~\ref{fig:14001_all} shows that, on MI-I, the CR augmented samples from S1 and S4 were far away from their original samples. This is because these two subjects had lower data quality, as indicated by the low classification accuracies in Table~4 and previous research \citep{Zanini2018RA}. Figure~\ref{fig:5a} also shows that their two classes had large overlaps. So, the CR approach may perform less well when the original data quality is low.
\item Figure~\ref{fig:mi1_sub} shows that, on MI-III, the distributions of CR augmented samples and the original samples from S0 and S5 were much separated than others. This is because S0 and S5 performed right hand versus feet MI tasks, whereas other subjects performed left hand versus right hand MI tasks.
\end{enumerate}

\subsection{Necessity of Symmetry in CR} \label{sect:symm}

Random shuffle (RS) was performed to verify the necessity of symmetry in CR. RS randomly exchanges the left and right hemisphere channels, without considering the channel symmetry. For example, in Figure~\ref{fig:CR}(a), FC3 may be switched with P2 in RS, whereas FC3 must be switched with FC4 in CR.

The results are shown in Table~\ref{tab:rs_cr}. CR always achieved much better performance, indicating the necessity of maintaining strict symmetry of the switched channels.

\begin{table*}[htpb]     \centering \setlength{\tabcolsep}{2mm} \renewcommand\arraystretch{1}
       \caption{Performance comparison of RS and CR in cross-subject unsupervised transfer. }  \label{tab:rs_cr}
    \begin{tabular}{cccccccccc}   \toprule
Approach & MI-I & MI-II & MI-III & SSVEP & ERP-I & ERP-II & Seizure-I & Seizure-II \\
\bottomrule
RS & 60.54$_{\pm0.45}$ & 48.69$_{\pm2.03}$ & 67.44$_{\pm0.80}$ & 76.25$_{\pm0.32}$ & 71.76$_{\pm0.18}$ & 79.50$_{\pm0.21}$ & 63.39$_{\pm0.23}$ & 75.25$_{\pm1.09}$ \\
CR & 75.77$_{\pm1.58}$ & 69.11$_{\pm0.84}$ & 74.97$_{\pm0.96}$ & 83.12$_{\pm0.17}$ & 72.71$_{\pm0.13}$ & 83.02$_{\pm0.11}$ & 66.04$_{\pm0.16}$ & 79.47$_{\pm0.57}$ \\
\bottomrule
\end{tabular}
\end{table*}

\subsection{Effect of Transfer Learning}

\begin{figure*}[htpb]\centering
\includegraphics[width=.9\linewidth,clip]{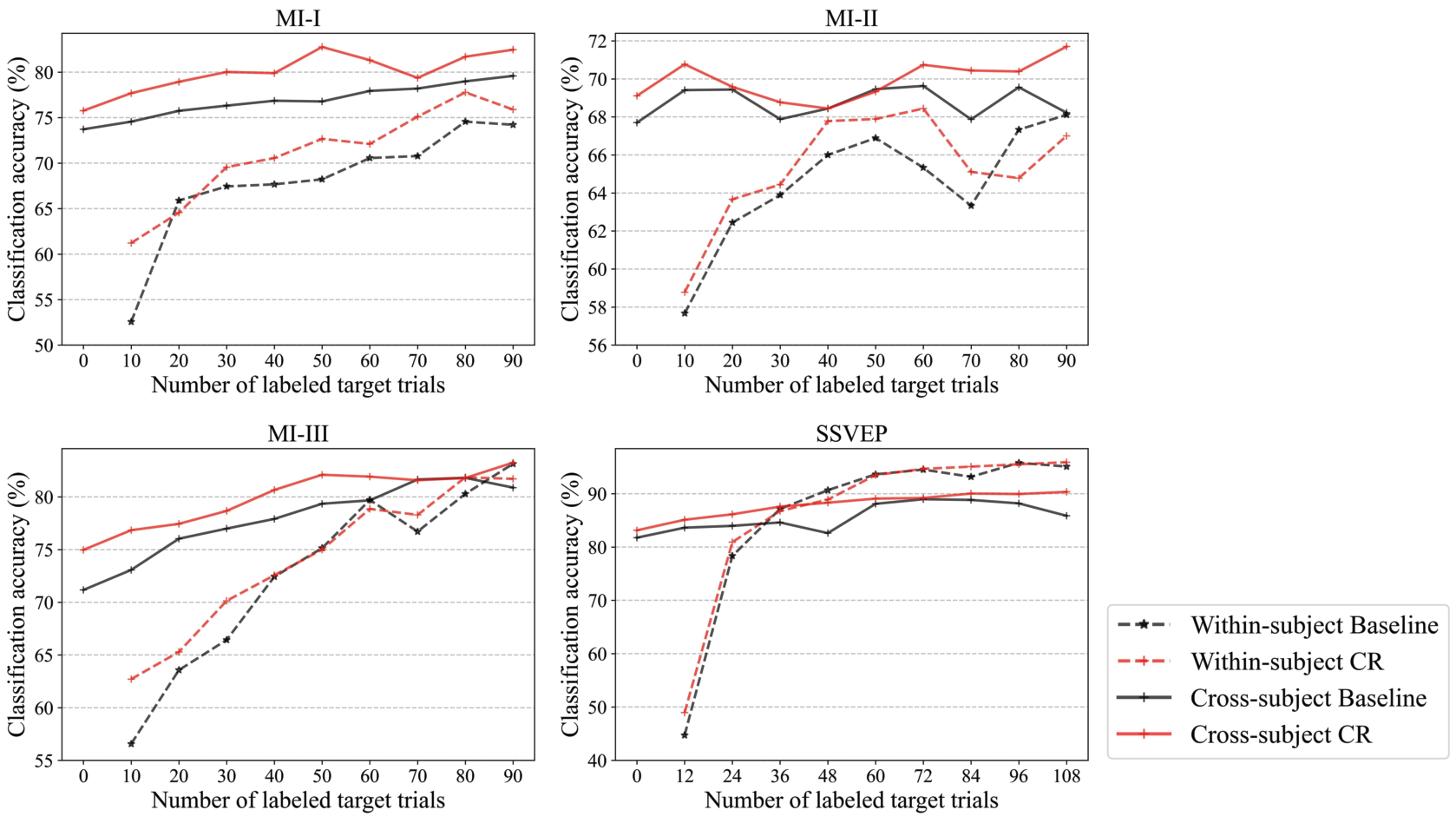}
\caption{Performance of CR as the number of labeled target samples increased on MI-I, MI-II, MI-III and SSVEP datasets.} \label{fig:nt_increasing}
\end{figure*}

\citet{Wu2022} demonstrated the benefits of utilizing source subjects' data to facilitate the calibration for the target subject. Figure~\ref{fig:nt_increasing} shows the performance of CR as the number of labeled target samples increased:
\begin{enumerate}
\item Generally, as the number of labeled target samples increased, the performance in both scenarios increased, regardless of whether data augmentation was used or not, which is intuitive.
\item Cross-subject supervised transfer almost always outperformed within-subject classification (the solid curves are almost always higher than the dashed curves of the same color), especially when the number of labeled target samples was small, indicating again the benefits of utilizing source subjects' data for target subject model learning.
\item As the number of labeled target samples increased, the performance improvement of transfer learning diminished. It indicates that if we have access to an adequate amount of labeled target data, then auxiliary data from other subjects are less beneficial. For example, on MI-I, when there were 10 labeled target samples, the performance improvement of cross-subject CR over within-subject CR was 77.69\%-61.22\%=16.47\%; however, when there were 90 labeled target samples, the performance improvement of cross-subject CR over within-subject CR reduced to 82.47\%-75.89\%=6.58\%.
\item CR almost always outperformed Baseline (the red curves are almost always higher than the corresponding black curves of the same line style), indicating again the effectiveness and robustness of incorporating prior knowledge in data augmentation.
\end{enumerate}

\begin{figure*}[htpb]\centering
\includegraphics[width=1\linewidth,clip]{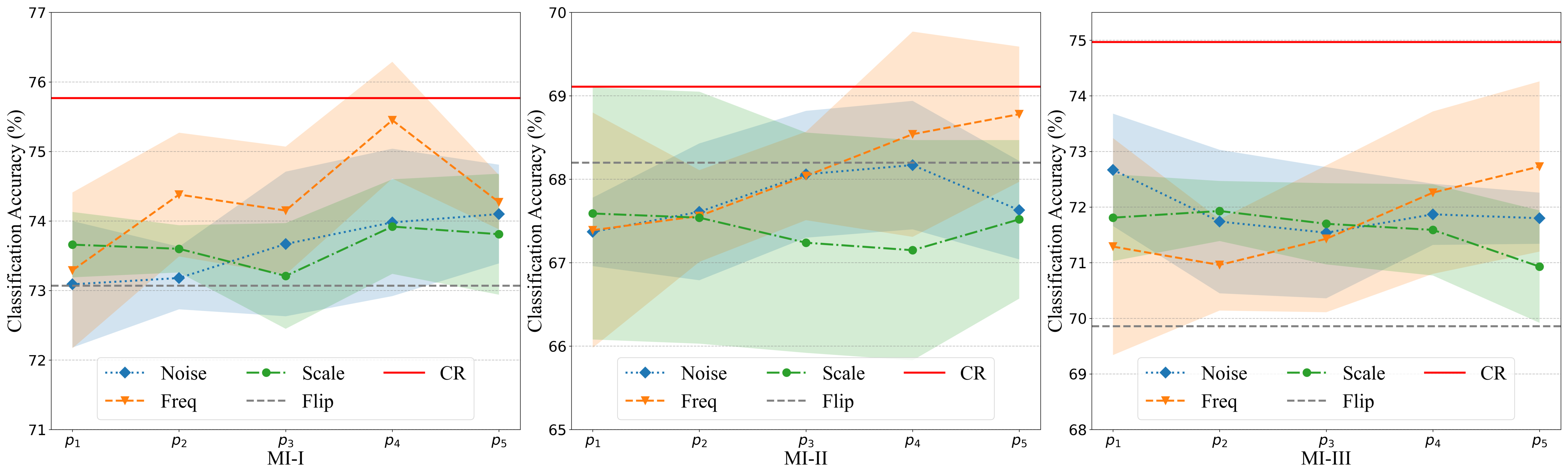}
\caption{Parameter sensitivity analysis for Noise, Freq, Scale and Flip on MI-I, MI-II and MI-III in cross-subject unsupervised transfer. $C_{\text{noise}}\in\{0.25,0.5,1,2,4\}$, $C_{\text{freq}}\in\{0.1,0.2,0.3,0.4,0.5\}$, and $C_{\text{scale}}\in\{0.005,0.01,0.05,0.1,0.2\}$. CR and Flip do not have hyperparameters. }\label{fig:para_sen}
\end{figure*}

\subsection{Hyperparameter Analysis} \label{sect:parameter}

CR is hyperparameter-free; however, other data augmentation approaches like Noise, Freq and Scale all have hyperparameters, which may affect their performance. Figure~\ref{fig:para_sen} shows how their performance changed with the hyperparameters. Although better performance of the other three data augmentation approaches may be obtained by adjusting their hyperparameters, our proposed CR almost always outperformed them.

\subsection{Combination of Different Data Augmentations}

It is also interesting to study if our proposed CR can be combined with other data augmentation approaches for further performance improvement.

Figure~\ref{fig:para_sen} shows that Freq has the second-best performance in MI-I when $C_{\text{freq}}=0.4$, and in MI-II and MI-III when $C_{\text{freq}}=0.5$. We combined these best configurations of Freq with CR, expanding the training data size by a factor of 6. The results are shown in Table~\ref{tab:mergeDA}. CR+Freq always outperformed Freq, and achieved comparable or better performance than CR, suggesting the effectiveness, robustness and flexibility of CR.

\begin{table}[htpb]     \centering \setlength{\tabcolsep}{2mm} \renewcommand\arraystretch{1}
       \caption{The performance of combining CR with Freq on MI datasets in cross-subject unsupervised transfer setting. The best average performance is marked in bold.}  \label{tab:mergeDA}
    \begin{tabular}{cccccccccc}   \toprule
Approach & MI-I & MI-II & MI-III \\
\bottomrule
Freq & 75.45$_{\pm0.89}$ & 68.78$_{\pm0.55}$ & 72.73$_{\pm0.82}$ \\
CR  & \textbf{75.77}$_{\pm1.58}$ & 69.11$_{\pm0.84}$ & 74.97$_{\pm0.96}$ \\
CR+Freq & 75.45$_{\pm0.62}$ & \textbf{69.70}$_{\pm0.73}$ & \textbf{77.27}$_{\pm0.84}$ \\
\bottomrule
\end{tabular}
\end{table}

\section{Conclusions}\label{sect:conclusions}

To cope with the typical calibration data shortage challenge in EEG-based BCIs, this paper has proposed a parameter-free CR data augmentation approach that incorporates prior knowledge on the channel distributions of different BCI paradigms in data augmentation. Experiments on eight public EEG datasets across four different BCI paradigms (MI, SSVEP, P300, and Seizure classifications) using different decoding algorithms demonstrated that: 1) CR is effective, i.e., it can noticeably improve the classification accuracy; 2) CR is robust, i.e., it consistently outperforms other data augmentation approaches in the literature; and, 3) CR is flexible, i.e., it can be combined with other data augmentation approaches to further increase the performance. We suggest that data augmentation approaches like CR should be an essential step in EEG signal classification.

CR also has some limitations, which will be investigated in our future research:
\begin{enumerate}
\item Although CR is effective for augmenting left/right hand MI trials, it cannot be directly applied to other classes like feet or tongue MIs.
\item This paper assumes strict symmetry between left and right hemisphere EEG electrodes; however, in practice the electrodes may not always be perfectly symmetric.
\end{enumerate}

\end{document}